\documentclass[showpacs]{revtex4}

\usepackage{graphicx}
\usepackage{wrapft}
\usepackage{amsmath,amssymb}
\usepackage{bm}
\usepackage{color}
\usepackage{textcomp}
\usepackage{multirow}

\begin{document}

\title{Cluster structures of excited states in $^{14}$C}

\author{Tadahiro Suhara and Yoshiko Kanada-En'yo}

\affiliation{Department of Physics, Graduate School of Science, Kyoto University, Kyoto 606-8502, Japan}

\begin{abstract}
Structures of excited states in $^{14}$C are investigated 
with a method of $\beta$-$\gamma$ constraint 
antisymmetrized molecular dynamics in combination with 
the generator coordinate method.
Various excited states with the developed $3\alpha$-cluster core structures 
are suggested in positive- and negative-parity states.
In the positive-parity states, triaxial deformed and linear-chain structures are found to construct excited bands.
Interestingly, $^{10}$Be+$\alpha$ correlation is found in the cluster states above the $^{10}$Be+$\alpha$ threshold energy.
\end{abstract}


\maketitle

\section{Introduction}\label{introduction}

Owing to the experimental progress of unstable nuclei, 
various exotic structures have been discovered in neutron-rich nuclei 
where excess neutrons play important roles. 
In light neutron-rich nuclei, cluster structure with excess neutrons is one of the hot subjects in experimental and theoretical studies, 
where searching for new cluster states in excited states has been performed.

In this point of view, neutron-rich C isotopes such as $^{14}$C are an interesting subject 
because they have excess neutrons compared with $^{12}$C, which is already well known to show various structures 
owing to the coexistence of shell-model and cluster features.
The ground state of $^{12}$C has mainly a shell-model feature of the $p_{3/2}$-subshell closed configuration, 
whereas, in the excited states above the 3$\alpha$ threshold energy, 
various configurations of 3$\alpha$-cluster structures were suggested in many theoretical works \cite{Horiuchi_OCM_74,Uegaki_12C_77,Kamimura_12C_77,Descouvemont_12C_87,En'yo_12C_98,Tohsaki_12C_01,Funaki_12C_03,Funaki_12C_05,Neff_12C_04,En'yo_12C_07,Kurokawa_12C_07}.
For example, the $3^{-}_{1}$ state has been discussed in association with an equilateral-triangular structure of three $\alpha$ clusters. 
The $\alpha$ condensation of weakly interacting three $\alpha$ clusters suggested in the $0^{+}_2$ and $2^{+}_{2}$ states 
is another cluster aspect attracting great interest recently \cite{Tohsaki_12C_01,Funaki_12C_03,Funaki_12C_05}.
Moreover, a linear-chainlike structure with an obtuse-angle-triangular $3\alpha$ configuration 
was suggested in the $0^{+}_{3}$ state \cite{Neff_12C_04,En'yo_12C_07}.
Therefore, it is expected that $^{14}$C with two excess neutrons may also exhibit rich phenomena in the excited states. 
In particular, from an analogy of $^{12}$C, various cluster structures may appear in $^{14}$C as well as shell-model structures.

For the excited states of $^{14}$C, there are many experimental studies 
that indicate the appearance of cluster states such as $^{10}$Be+$\alpha$ cluster states
\cite{vonOertzen_14C_04,Milin_14C_04,Soic_CandBisotopes_04,Price_14C_07,Haigh_14C_08}.
Also on the theoretical side, cluster structures of $^{14}$C were suggested \cite{Itagaki_14C_04}. 
In the study of $^{14}$C with $3\alpha$+$2n$ cluster models, Itagaki \textit{et al.} predicted that
an equilateral-triangular structure of the well-developed three $\alpha$ clusters surrounded by excess neutrons 
is formed constructing $K^\pi=0^{+}$ and $K^\pi=3^{-}$ rotational bands in the excited states.
They argued that these two bands originate in a rather rigid $3\alpha$ structure 
stabilized by the excess neutrons in the molecular orbitals. 
This mechanism can be interpreted as a realization of the $\alpha$ crystallization in dilute nuclear medium.
Another interesting problem is whether a linear-chain $3\alpha$ structure with two excess neutrons exists in the excited states of $^{14}$C. 
Although there were discussions on the possibility of the linear-chain structure 
in neutron-rich C isotopes \cite{vonOertzen_Cisotopes_97,Itagaki_Cisotopes_01,vonOertzen_14C_04}, 
there is no clear conclusion for stability of the linear-chain structure in $^{14}$C.
Thus, the excited states of $^{14}$C are attracting much interest recently, 
and therefore, systematic study of the ground and excited states is required.

Our aim is to investigate the ground and excited states of $^{14}$C while focusing on cluster features. 
To clarify the role of excess neutrons in neutron-rich nuclei, 
it is helpful to consider analogies and differences of cluster features between $^{14}$C and $^{12}$C.
It is also a challenging problem to search for new types of structures in $^{14}$C.
In such a systematic study, it is rather important to apply a framework which is free from model assumptions of clusters
because the formation of three $\alpha$ clusters is not obvious in $^{14}$C.
Therefore, we adopted a method of antisymmetrized molecular dynamics (AMD), 
which is a framework without any assumption of the existence of clusters.
In the present study, we apply a combination of the $\beta$-$\gamma$ constraint AMD and the generator coordinate method (GCM), 
which we call the $\beta$-$\gamma$ constraint AMD + GCM.
This method has already been proved to be a powerful approach to describe various structures
such as cluster and shell-model-like structures \cite{Suhara_AMD_10}. 
In particular, it is useful for the systematic study of cluster states in excited states 
owing to the superposition of basis AMD wave functions on the two-dimensional $\beta$-$\gamma$ plane. 
Therefore, it is suitable for the study of $^{14}$C.

This article is organized as follows. 
In Sec. \ref{framework}, we explain the framework of the $\beta$-$\gamma$ constraint AMD + GCM.
The calculated results are shown in Sec. \ref{results}.
In Sec. \ref{discussion}, we discuss the effect of excess neutrons and compare the present results with earlier works.
Finally, in Sec. \ref{summary}, a summary and an outlook are given.

\section{Framework of the $\beta$-$\gamma$ constraint AMD + GCM}\label{framework}

The frameworks of AMD are described in detail, for example, in Refs.~\cite{En'yo_PTP_95,En'yo_AMD_95,En'yo_sup_01,En'yo_AMD_03}.
In this article, we adopt a version of AMD, the $\beta$-$\gamma$ constraint AMD \cite{Suhara_AMD_10}, 
in which we perform the variation with the constraint on the quadrupole deformation parameters, $\beta$ and $\gamma$.

\subsection{Wave function of AMD}

In the method of AMD, 
a basis wave function of an $A$-nucleon system $|\Phi \rangle$ 
is described by a Slater determinant of single-particle wave functions $|\varphi_{i} \rangle$ as
\begin{equation}
|\Phi \rangle = \frac{1}{\sqrt{A!}} \det \left\{ |\varphi_{1} \rangle, \cdots ,|\varphi_{A} \rangle \right\}.
\end{equation}
The $i$-th single-particle wave function $|\varphi_{i} \rangle$ consists of 
the spatial part $|\phi_{i} \rangle$, spin part $|\chi_{i} \rangle$, and isospin part $|\tau_{i} \rangle$ as
\begin{equation}
	|\varphi_{i} \rangle = |\phi_{i} \rangle |\chi_{i} \rangle |\tau_{i} \rangle.
\end{equation}
The spatial part $|\phi_{i} \rangle$ is given by a Gaussian wave packet
whose center is located at $\bm{Z}_{i}/\sqrt{\nu}$ as
\begin{equation}
	\langle \bm{r} | \phi_{i} \rangle = \left( \frac{2\nu}{\pi} \right)^{\frac{3}{4}}
		\exp \left[ - \nu \left( \bm{r} - \frac{\bm{Z}_{i}}{\sqrt{\nu}} \right)^{2} 
		+ \frac{1}{2} \bm{Z}_{i}^{2}\right] 
	\label{single_particle_spatial}, 
\end{equation}
where $\nu$ is the width parameter and is taken to be a common value for all the
single-particle Gaussian wave functions in the present work.
The spin orientation is given by the parameter $\bm{\xi}_{i}$, while
the isospin part $|\tau_{i} \rangle$ is fixed to be up (proton) or down (neutron), 
\begin{align}
	|\chi_{i} \rangle &= \xi_{i\uparrow} |\uparrow \ \rangle + \xi_{i\downarrow} |\downarrow \ \rangle,\\
	|\tau_{i} \rangle &= |p \rangle \ \text{or} \ |n \rangle.
\end{align}
In a basis wave function $|\Phi \rangle$, $\{ X \} \equiv \{ \bm{Z} , \bm{\xi} \} = \{ \bm{Z}_{1} , \bm{\xi}_{1} , \bm{Z}_{2} , \bm{\xi}_{2} , 
\cdots , \bm{Z}_{A} , \bm{\xi}_{A} \}$ are complex variational parameters and they 
are determined by the energy optimization.

\subsection{Parity and angular momentum projections}

We project the AMD wave function onto parity and angular momentum eigenstates
by using the parity projection operator $\hat{P}^{\pm}$ and the angular-momentum projection operator $\hat{P}^{J}_{MK}$.
The parity projection operator $\hat{P}^{\pm}$ is defined as 
\begin{equation}
	\hat{P}^{\pm} \equiv \frac{1 \pm \hat{P}}{2},
\end{equation}
where $\hat{P}$ is the parity operator.
The angular-momentum projection operator $\hat{P}^{J}_{MK}$ is defined as
\begin{equation}
	\hat{P}^{J}_{MK} \equiv \frac{2 J + 1}{8 \pi^{2}} \int d \Omega D^{J*}_{MK}(\Omega) \hat{R}(\Omega),
\end{equation}
where $\Omega = (\alpha, \beta, \gamma)$ are the Euler angles, $D^{J}_{MK}(\Omega)$ is the Wigner's $D$ function, 
and $\hat{R}(\Omega) = e^{- i \alpha \hat{J}_z} e^{- i \beta \hat{J}_y} e^{- i \gamma \hat{J}_z}$ is the rotation operator.

We perform the variation for the parity projected wave function $|\Phi ^{\pm} \rangle$ defined as
\begin{equation}
	|\Phi ^{\pm} \rangle \equiv P^{\pm} |\Phi \rangle.
\end{equation}
After the variation, we project the obtained wave function onto the total-angular-momentum eigenstates.
It means that the parity projection is performed before the variation, and
the total-angular-momentum projection is carried after the variation.

\subsection{Constraint variation on quadrupole deformation}

To describe various cluster and shell-model structures which may appear 
in the ground and excited states of $^{14}$C,
we perform the energy variation with the constraints on the quadrupole deformation parameters, $\beta$ and $\gamma$.

The deformation parameters, $\beta$ and $\gamma$, are defined as
\begin{align}
	&\beta \cos \gamma \equiv \frac{\sqrt{5\pi}}{3} 
		\frac{2\langle \hat{z}^{2} \rangle -\langle \hat{x}^{2} \rangle -\langle \hat{y}^{2} \rangle }{R^{2}}, \\
	&\beta \sin \gamma \equiv \sqrt{\frac{5\pi}{3}} 
		\frac{\langle \hat{x}^{2} \rangle -\langle \hat{y}^{2} \rangle }{R^{2}} \label{definition_beta_gamma}, \\
	&R^{2} \equiv \frac{5}{3} \left( \langle \hat{x}^{2} \rangle + \langle \hat{y}^{2} \rangle 
		+ \langle \hat{z}^{2} \rangle \right).
\end{align}
Here, $\langle \hat{O} \rangle$ represents the expectation value of the operator $\hat{O}$ for an intrinsic wave function $| \Phi \rangle$.
$\hat{x}$, $\hat{y}$, and $\hat{z}$ are the inertia principal axes that are chosen as
$\langle \hat{y}^{2} \rangle \le \langle \hat{x}^{2} \rangle \le \langle \hat{z}^{2} \rangle $ and
$\langle \hat{x} \hat{y} \rangle = \langle \hat{y} \hat{z} \rangle = \langle \hat{z} \hat{x} \rangle =0$.
To satisfy the latter condition, we also impose 
the constraints $\langle \hat{x} \hat{y} \rangle/R^{2} = \langle \hat{y} \hat{z} \rangle/R^{2} = \langle \hat{z} \hat{x} \rangle/R^{2}  = 0$. 
To obtain the energy minimum state under the constraint condition,
we add the constraint potential $V_{\text{const}}$ to the total energy of the system
in the energy variation. The constraint potential $V_{\text{const}}$ is given as
\begin{align} 
	V_{\text{const}} \equiv &\eta_{1} 
	\left[ (\beta \cos \gamma - \beta_{0} \cos \gamma_{0})^{2} + (\beta \sin \gamma - \beta_{0} \sin \gamma_{0})^{2} \right] \notag \\
	+ &\eta_{2} \left[ \left( \frac{\langle \hat{x} \hat{y} \rangle}{R^{2}} \right)^{2} 
		+ \left( \frac{\langle \hat{y} \hat{z} \rangle}{R^{2}} \right)^{2} 
		+ \left( \frac{\langle \hat{z} \hat{x} \rangle}{R^{2}} \right)^{2} \right].
	\label{constraint_energy}
\end{align}
Here, $\eta_{1}$ and $\eta_{2}$ take sufficiently large values.
That is, we minimize the constrained energy $E^{\pm}_{\text{const}}$ defined as
\begin{equation}
	E^{\pm}_{\text{const}} \equiv \frac{\langle \Phi^{\pm} | \hat{H} | \Phi^{\pm} \rangle}{\langle \Phi^{\pm} | \Phi^{\pm} \rangle} + V_{\text{const}},
\end{equation}
where $\hat{H}$ is the Hamiltonian, to determine $\{ X \}$.
After the variation with the constraints, we obtain the optimized wave functions
$|\Phi^{\pm}(\beta_{0}, \gamma_{0}) \rangle$
for each set of parameters, $(\beta, \gamma) = (\beta_{0}, \gamma_{0})$.

\subsection{Generator coordinate method}

In the calculations of energy levels, 
we superpose the parity and total-angular-momentum projected 
AMD wave functions $\hat{P}^{J}_{MK} |\Phi^{\pm}(\beta, \gamma) \rangle$. 
Thus, the final wave function for the $J^\pm_n$ state is given by
a linear combination of the basis wave functions as 
\begin{equation}
	|\Phi ^{J\pm}_{n} \rangle = \sum_{K} \sum_{i} f_{n}(\beta_{i}, \gamma_{i}, K) \hat{P}^{J}_{MK} |\Phi^{\pm}(\beta_{i}, \gamma_{i}) \rangle.
	\label{dispersed_GCM}
\end{equation}
The coefficients $f_{n}(\beta_{i}, \gamma_{i}, K)$ are determined using the Hill-Wheeler equation
\begin{equation}
	\delta \left( \langle \Phi ^{J\pm}_{n} | \hat{H} | \Phi ^{J\pm}_{n} \rangle - 
	E_{n} \langle \Phi ^{J\pm}_{n} | \Phi ^{J\pm}_{n} \rangle\right) = 0.
	\label{Hill-Wheeler}
\end{equation}
This means the superposition of multiconfigurations described by 
parity and total-angular-momentum projected AMD wave functions.
In the limit of sufficient basis wave functions 
on the $\beta$-$\gamma$ plane, it corresponds to the
GCM with the two-dimensional generator coordinates 
of the quadrupole deformation parameters, $\beta$ and $\gamma$.

\subsection{Hamiltonian and parameters}

The Hamiltonian $\hat{H}$ consists of the kinetic term 
and effective two-body interactions as
\begin{equation}
	\hat{H} = \sum_{i} \hat{t}_{i} - \hat{T}_{\text{G}} + \sum_{i<j} \hat{V}^{\text{central}}_{ij} 
	+ \sum_{i<j} \hat{V}^{\text{spin-orbit}}_{ij} + \sum_{i<j} \hat{V}^{\text{Coulomb}}_{ij},
\end{equation}
where $\hat{V}^{\text{central}}_{ij}$, $\hat{V}^{\text{spin-orbit}}_{ij}$, and $\hat{V}^{\text{Coulomb}}_{ij}$ 
are the central force, spin-orbit force, and Coulomb force, respectively.
As the central force, we use the Volkov No.~2 interaction \cite{Volkov_No2_65},
\begin{equation}
	\hat{V}^{\text{central}}_{ij} = \sum_{k=1}^{2} v_{k} \exp \left[- \left( \frac{\hat{r}_{ij}}{a_{k}} \right)^{2} \right]
                                  (W + B \hat{P}_{\sigma} - H \hat{P}_{\tau} - M \hat{P}_{\sigma} \hat{P}_{\tau}),
\end{equation}
where $v_{1} = -60.65$ MeV, $v_{2} = 61.14$ MeV, $a_{1} = 1.80$ fm, and $a_{2} = 1.01$ fm.
For the spin-orbit part, we used the spin-orbit term of the G3RS interaction \cite{G3RS_79}, which is a two-range
Gaussian with a projection operator $\hat{P}(^{3} \mbox{O})$ onto the triplet odd state,
\begin{align}
	\hat{V}^{\text{spin-orbit}}_{ij} = \sum_{k=1}^{2} &u_{k} \exp \left[- \left( \frac{\hat{r}_{ij}}{b_{k}} \right)^{2} \right] \hat{P}(^{3} \mbox{O}) \hat{\bm{L}} \cdot \hat{\bm{S}}, \\
	\hat{P}(^{3} \mbox{O}) &= \frac{1+\hat{P}_{\sigma}}{2} \frac{1+\hat{P}_{\tau}}{2},
\end{align}
where $b_{1} = 0.600$ fm and $b_{2} = 0.477$ fm.
Here, $\hat{P}_{\sigma}$ and $\hat{P}_{\tau}$ are the spin and isospin exchange operators, respectively.

We take the same interaction parameters as those in Ref. \cite{Suhara_AMD_10}, i.e.,
the Majorana exchange parameter $M = 0.6$ ($W = 0.4$), the Bartlett exchange parameter $B = 0.125$,
and the Heisenberg exchange parameter $H = 0.125$ in the central force, and 
$u_{1} = -1600$ MeV and $u_{2} = 1600$ MeV in the spin-orbit force.
All these parameters are 
the same as those adopted in the studies for $^{9}$Be \cite{Okabe_parameter_79}, $^{10}$Be \cite{Itagaki_10Be_00}, 
and C isotopes, such as $^{12}$C, $^{14}$C, and $^{16}$C \cite{Itagaki_Cisotopes_01,Itagaki_14C_04}, except for a small modification in the strength of the spin-orbit force.
They are adjusted to reproduce 
the $\alpha + \alpha$ phase shift ($M$, $W=1-M$, $a_{1}$, $a_{2}$), binding energy of the deuteron ($B=H$),
and $\alpha + n$ phase shift ($u_{1}=-u_{2}$, $b_{1}$, $b_{2}$).
We adopt the slightly weaker strengths of the spin-orbit force 
than $-u_{1} = u_{2} = 2000$ MeV adopted in Refs.~\cite{Okabe_parameter_79,Itagaki_10Be_00,Itagaki_Cisotopes_01,Itagaki_14C_04}
to fit the $0^{+}_{1}$ energy of $^{12}$C \cite{Suhara_AMD_10}.

For the width parameter of single-particle Gaussian wave packets in Eq.~\eqref{single_particle_spatial}, 
we used the value $\nu = 0.235$ fm$^{-2}$, which is 
determined from a variational calculation 
for the ground state of $^{9}$Be in Ref.~\cite{Okabe_parameter_79}.
This value is also the same as those in the studies for Be isotopes \cite{Itagaki_10Be_00} 
and C isotopes \cite{Suhara_AMD_10,Itagaki_Cisotopes_01,Itagaki_14C_04}.

\section{Results}\label{results}

We applied the $\beta$-$\gamma$ constraint AMD + GCM to $^{14}$C.
In this section, we show the calculated results.

\subsection{Energy surfaces}

\begin{figure}[tb]
	\begin{tabular}{c}
	\includegraphics[width=8.6cm, bb=10 12 700 364, clip]{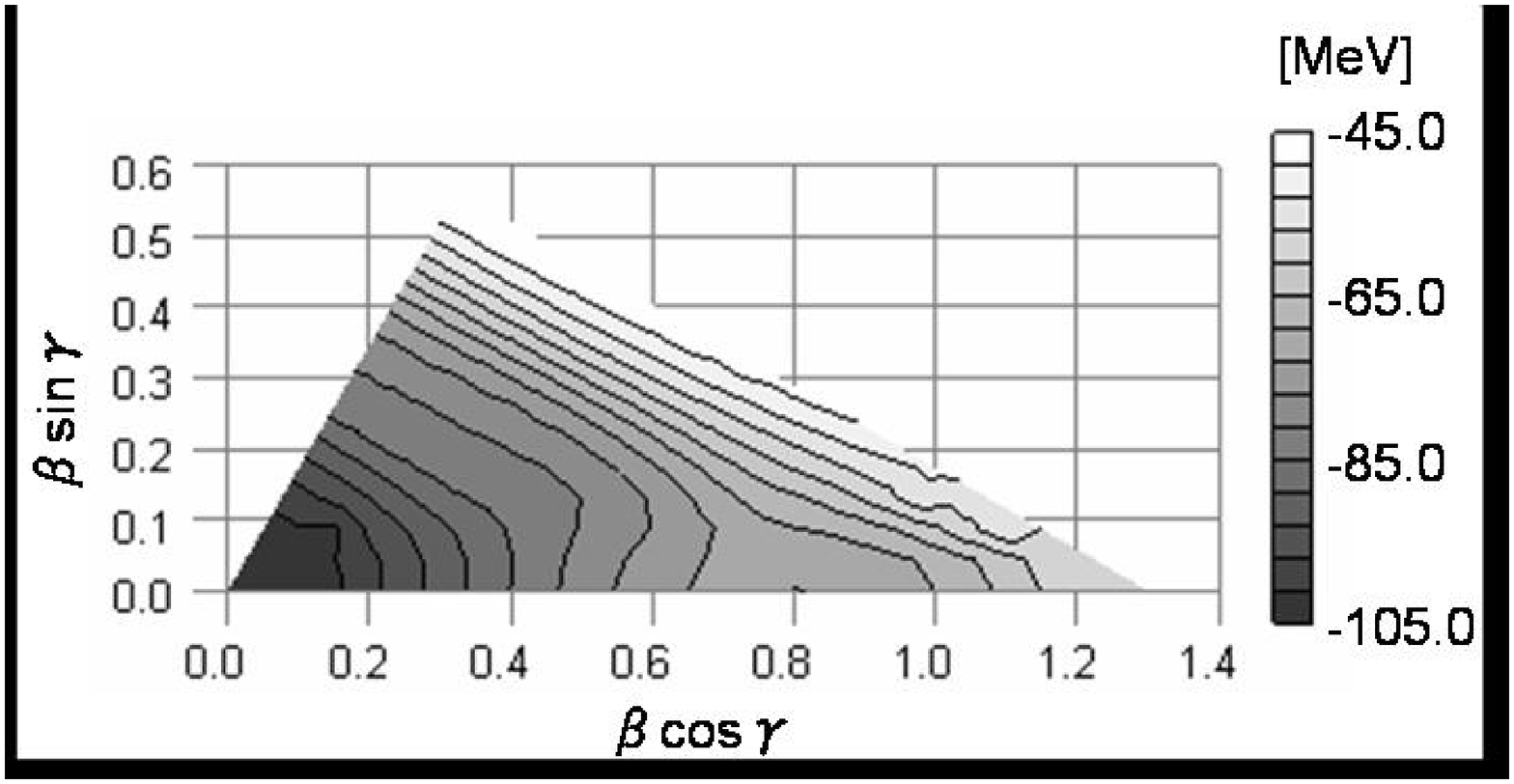} \\
	\includegraphics[width=8.6cm, bb=10 12 700 364, clip]{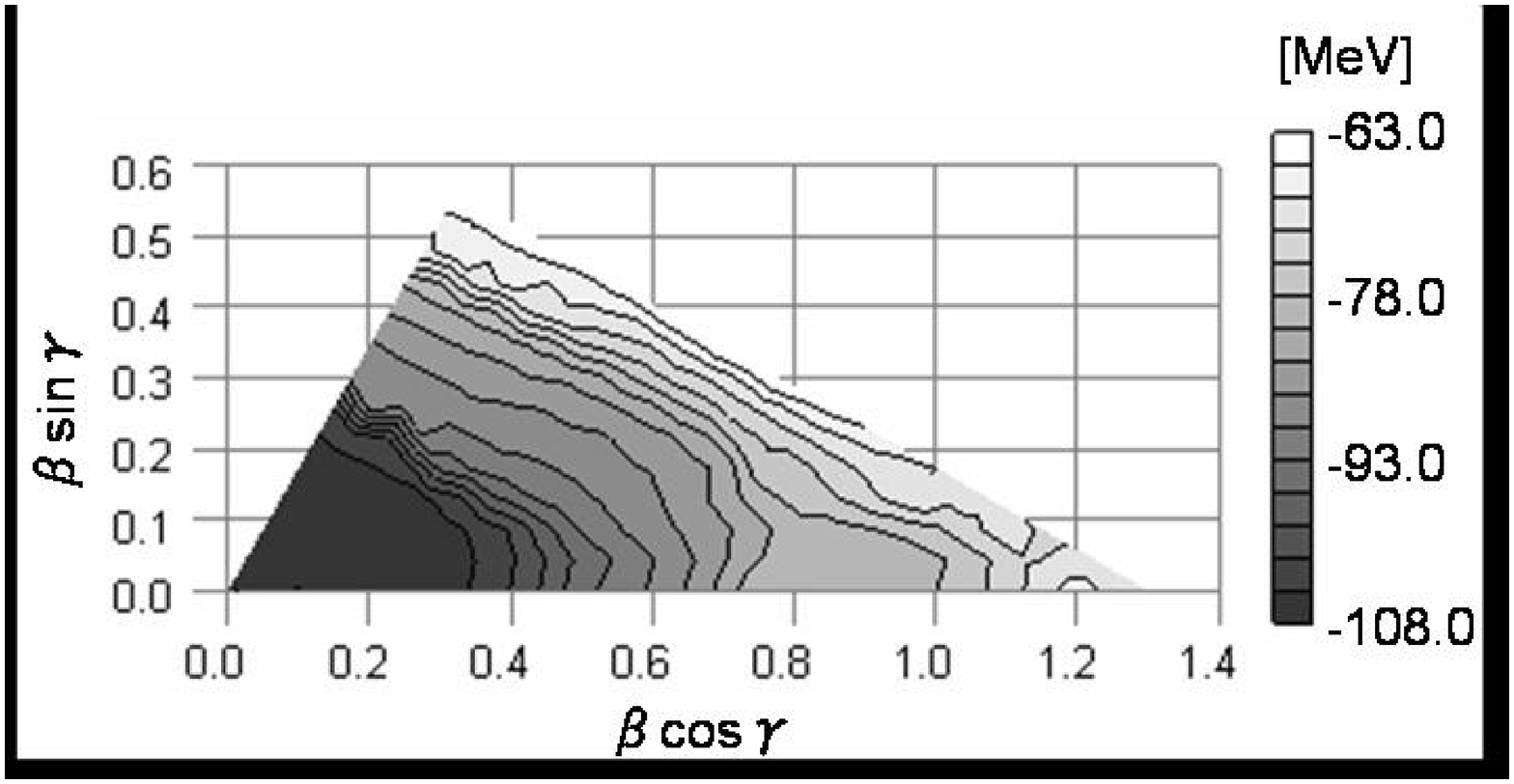} \\
	\end{tabular}
	\caption{Energy surfaces of $^{14}$C on the $\beta$-$\gamma$ plane.
	The upper shows the energy for the positive-parity states before the total-angular-momentum projection and
	the lower shows that for the $0^{+}$ states after the total-angular-momentum projection.}
	\label{14C_energy_surface_+}
\end{figure}

\begin{figure}[tb]
	\begin{tabular}{c}
	\includegraphics[width=8.6cm, bb=10 12 700 364, clip]{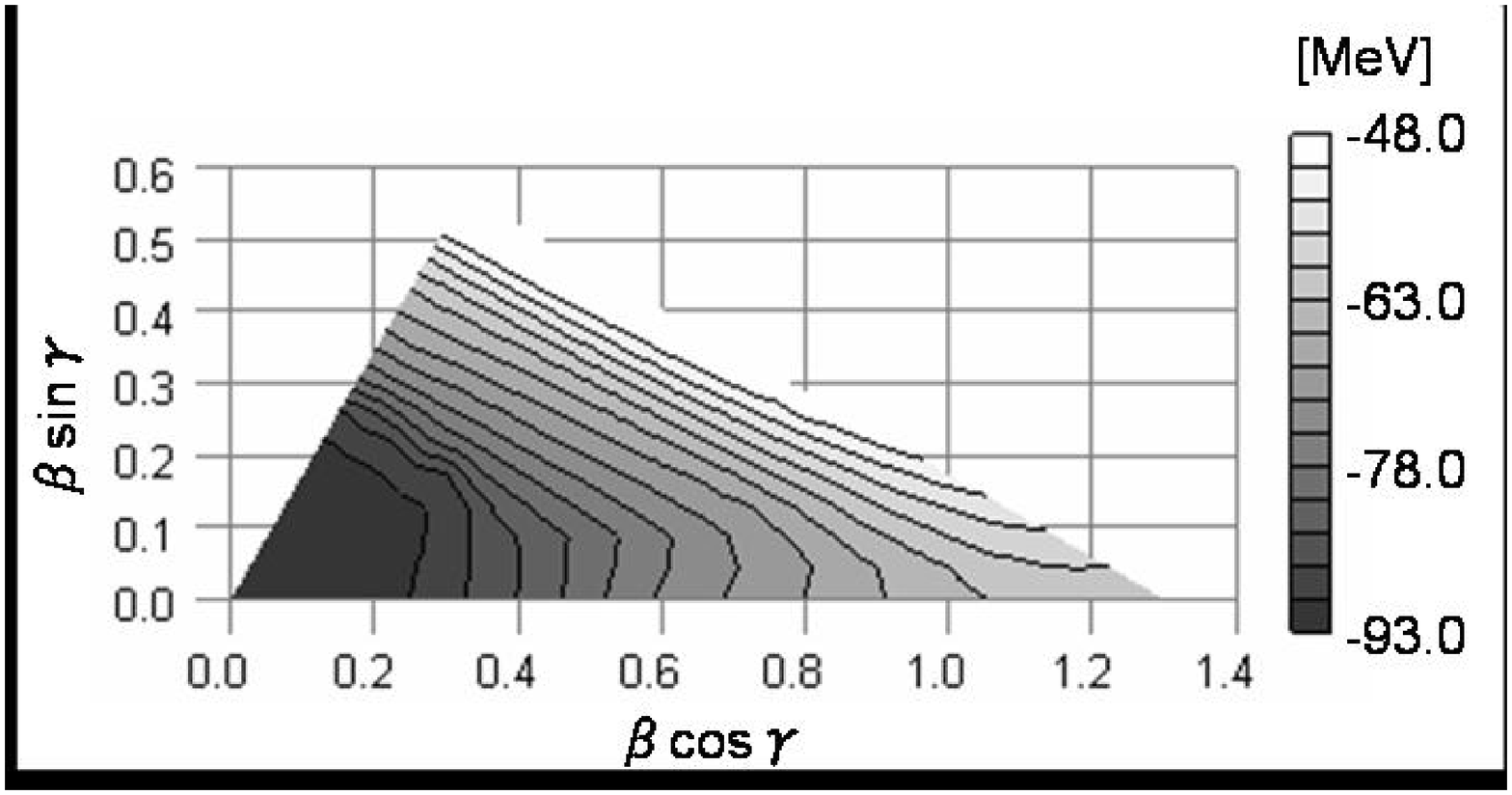} \\
	\includegraphics[width=8.6cm, bb=10 12 700 364, clip]{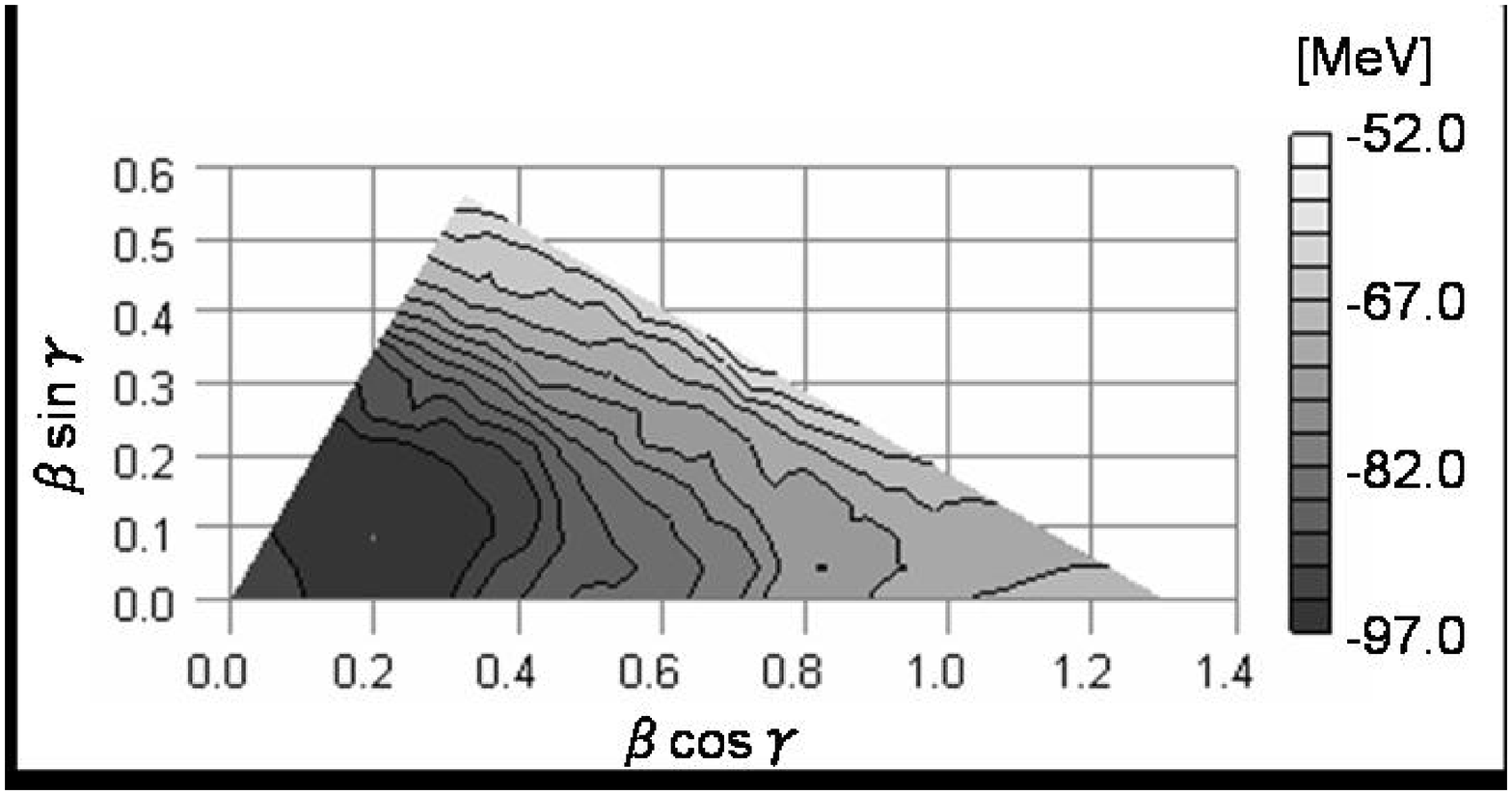} \\
	\end{tabular}
	\caption{Energy surfaces of $^{14}$C on the $\beta$-$\gamma$ plane.
	The upper shows the energy for the negative-parity states before the total-angular-momentum projection and
	the lower shows that for the $3^{-}$ states after the total-angular-momentum projection.}
	\label{14C_energy_surface_-}
\end{figure}

We performed variational calculations with the $\beta$-$\gamma$ constraint
at 196 mesh points of the triangle lattice on the $\beta$-$\gamma$ plane.
Energy surfaces as functions of $\beta$ and $\gamma$ are obtained.
The calculated energy surfaces on the $\beta$-$\gamma$ plane
for positive-parity states and negative-parity states are shown 
in Figs.~\ref{14C_energy_surface_+} and \ref{14C_energy_surface_-}, respectively.

In Fig.~\ref{14C_energy_surface_+}, the top panel shows the energy of 
the positive-parity states before the total-angular-momentum projection,
and the bottom panel shows the results for the $0^{+}$ states calculated 
by the total-angular-momentum projection after the variation.
We call the former the positive-parity energy surface and the latter the $0^{+}$ energy surface.
The minimum point of the positive-parity energy surface is at
$(\beta \cos \gamma, \beta \sin \gamma) = (0.00, 0.00)$, which indicates a spherical shape.
After the total-angular-momentum projection, the minimum point of the $0^{+}$ energy surface becomes
$(\beta \cos \gamma, \beta \sin \gamma) = (0.23, 0.04)$. 
It means that the deformation of the energy minimum state changes 
from the spherical shape before the total-angular-momentum projection 
to the triaxial shape after the projection, 
because higher correlations beyond mean field are incorporated 
by the total-angular-momentum projection.
In a largely deformed region, there exists a flat region around 
$(\beta \cos \gamma, \beta \sin \gamma) = (0.80 - 0.90,0.00)$ 
in the positive-parity energy surface. 
After the total-angular-momentum projection, 
a local minimum at $(\beta \cos \gamma, \beta \sin \gamma) = (0.90, 0.04)$ emerges from this flat region.
As we show later, a rotational band with the large prolate deformation 
is constructed by wave functions in this region after the GCM calculation.

The negative-parity energy surface and the $3^{-}$ energy surface are displayed in the 
top and bottom panels of Fig.~\ref{14C_energy_surface_-}, respectively.
The minimum point of the negative-parity energy surface is at
$(\beta \cos \gamma, \beta \sin \gamma) = (0.08, 0.04)$, 
and that of the $3^{-}$ energy surface is at $(\beta \cos \gamma, \beta \sin \gamma) = (0.20, 0.09)$. 
Again, the deformation of the energy minimum state changes 
from an almost spherical shape to a triaxial one before and 
after the total-angular momentum projection. 
In the largely deformed prolate region, 
the behavior of the energy surface for the negative-parity states 
is different from that of the positive-parity states.
Along the $\gamma = 0^{\circ}$ line, the energy rapidly increases 
and there is no flat region in the negative-parity surface.
Even after the total-angular-momentum projection, 
no flat region is found around the largely deformed prolate region as seen in the $3^{-}$ energy surface.

\subsection{Structures on the $\beta$-$\gamma$ plane}

In this section, we describe intrinsic structures obtained by the $\beta$-$\gamma$ constraint AMD.

We analyze the spatial configurations of the Gaussian centers $\{\bm{Z}_1,\bm{Z}_2,\cdots,\bm{Z}_A\}$ 
and the distributions of proton density $\rho_{p}$ and neutron density $\rho_{n}$ 
of each intrinsic wave function $|\Phi(\beta,\gamma) \rangle$.
We also investigate the neutron-proton density difference $\rho_{n} - \rho_{p}$ 
to observe excess neutron behaviors.
To demonstrate density distributions, we show the density $\Tilde{\rho}$ integrated along the $y$-axis as
\begin{align}
	& \Tilde{\rho} (x, z) \equiv \int dy \rho (\bm{r}), \\ 
	& \rho (\bm{r}) \equiv \langle \Phi (\beta, \gamma) | \sum_{i} \delta (\bm{r} - \hat{\bm{r}}_{i}) |\Phi (\beta, \gamma) \rangle.
	\label{density_distribution}
\end{align}

\begin{figure}[tb]
	\centering
	{\tabcolsep=0.7mm
	\begin{tabular}{ccccc}
	\vspace{0.3cm}
	 &
	\hspace{-0.2cm} $\Tilde{\rho}_{p}$ & \hspace{-0.2cm} $\Tilde{\rho}_{n}$ 
	& \hspace{-0.15cm}  $\Tilde{\rho}_{n} - \Tilde{\rho}_{p}$ & 
	\multirow{4}{*}{\begin{minipage}{1.8cm}[1/fm$^{2}$] \includegraphics[width=1.8cm, bb=379 301 450 490, clip]{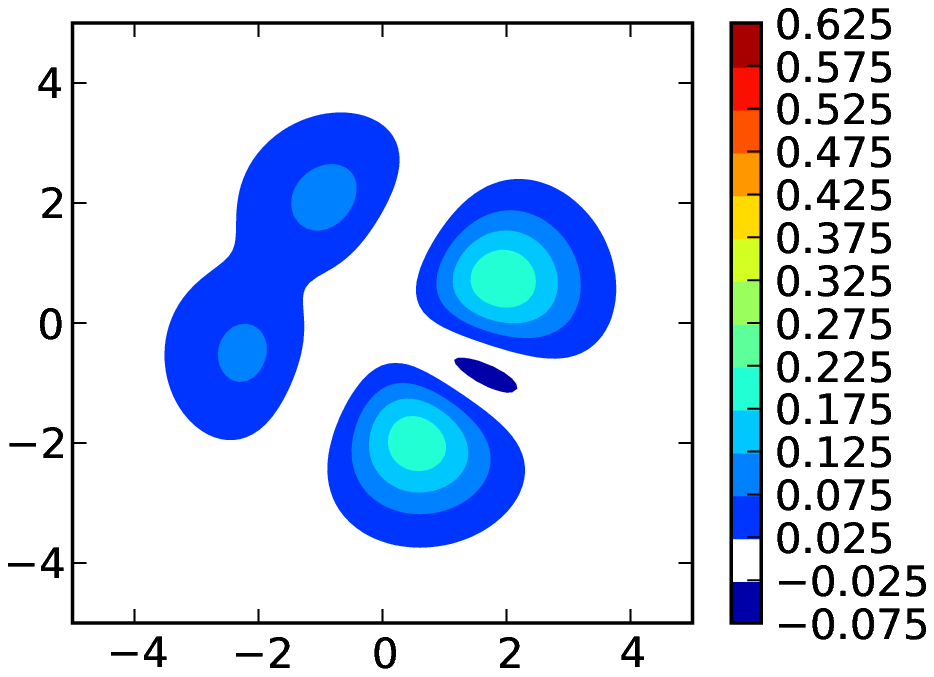}\end{minipage}} \\
	\begin{minipage}{0.9cm}\vspace{-1.7cm}(a) \end{minipage} &
	\includegraphics[width=1.8cm, bb=223 311 394 481, clip]{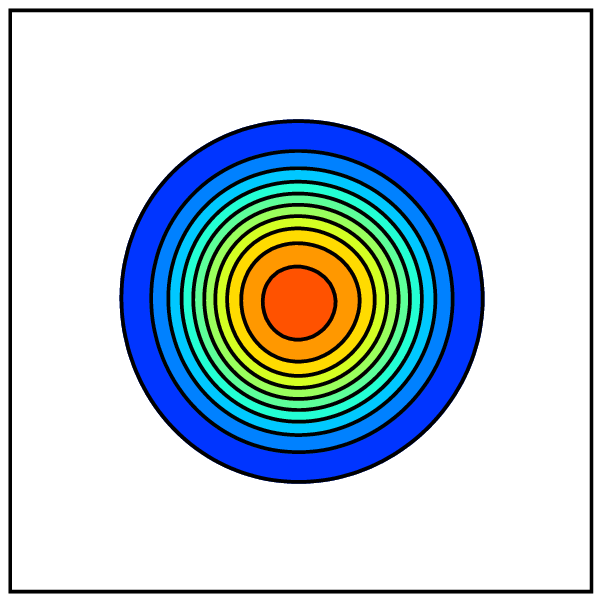} &
	\includegraphics[width=1.8cm, bb=223 311 394 481, clip]{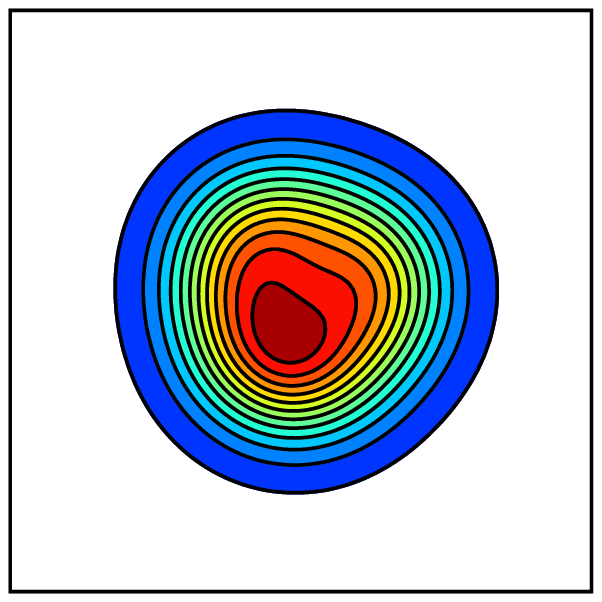} &
	\includegraphics[width=1.8cm, bb=223 311 394 481, clip]{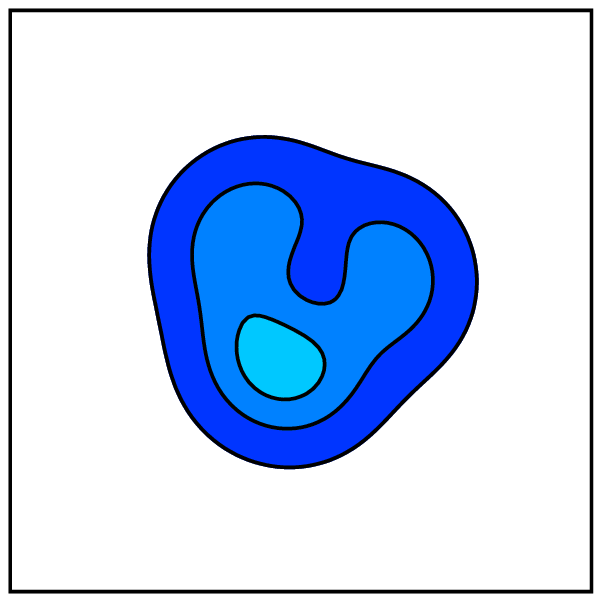} & \\
	\begin{minipage}{0.9cm}\vspace{-1.7cm}(b) \end{minipage} &
	\includegraphics[width=1.8cm, bb=223 311 394 481, clip]{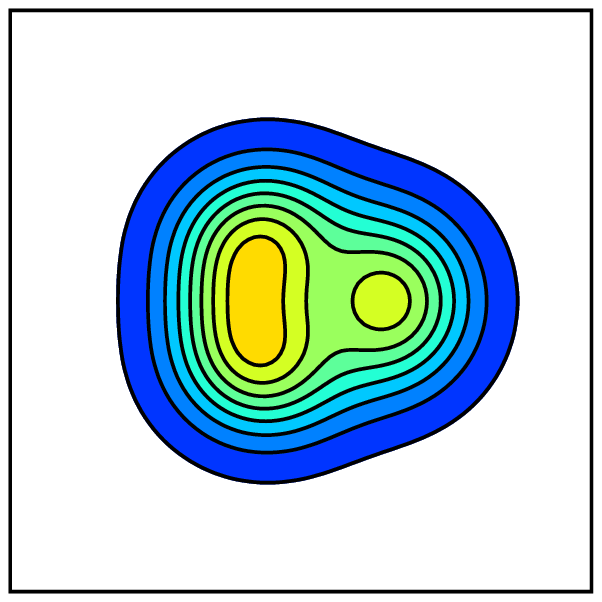} &
	\includegraphics[width=1.8cm, bb=223 311 394 481, clip]{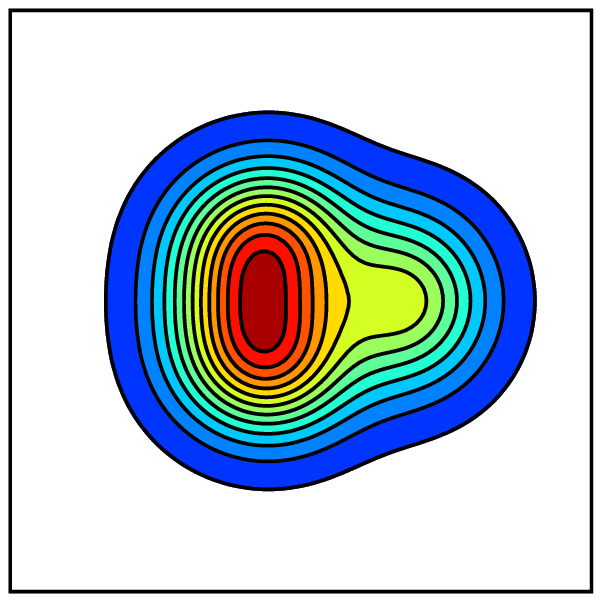} &
	\includegraphics[width=1.8cm, bb=223 311 394 481, clip]{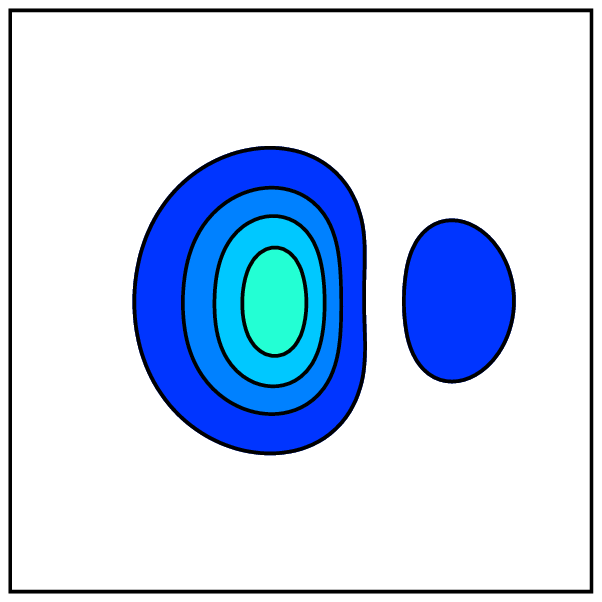} & \\
	\begin{minipage}{0.9cm}\vspace{-1.7cm}(c) \end{minipage} &
	\includegraphics[width=1.8cm, bb=223 311 394 481, clip]{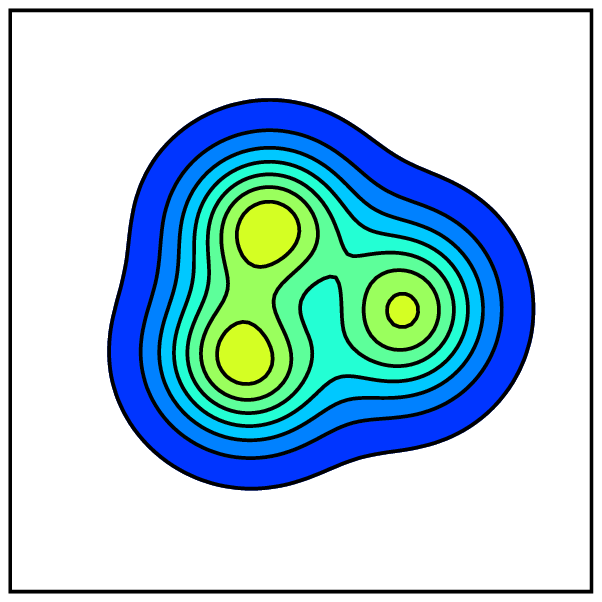} &
	\includegraphics[width=1.8cm, bb=223 311 394 481, clip]{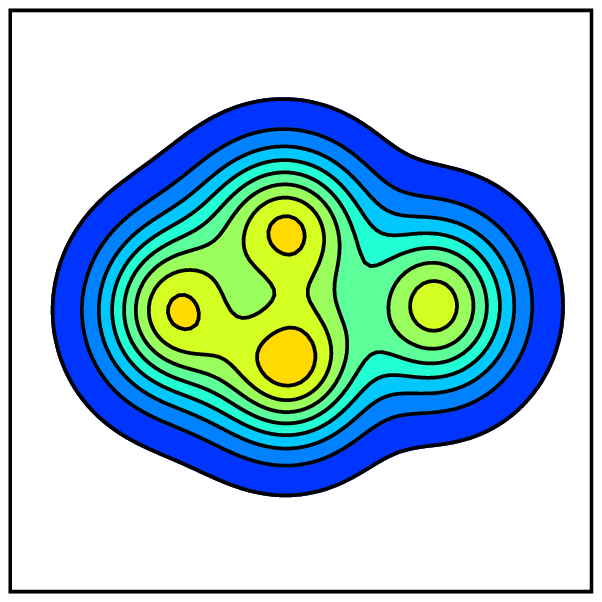} &
	\includegraphics[width=1.8cm, bb=223 311 394 481, clip]{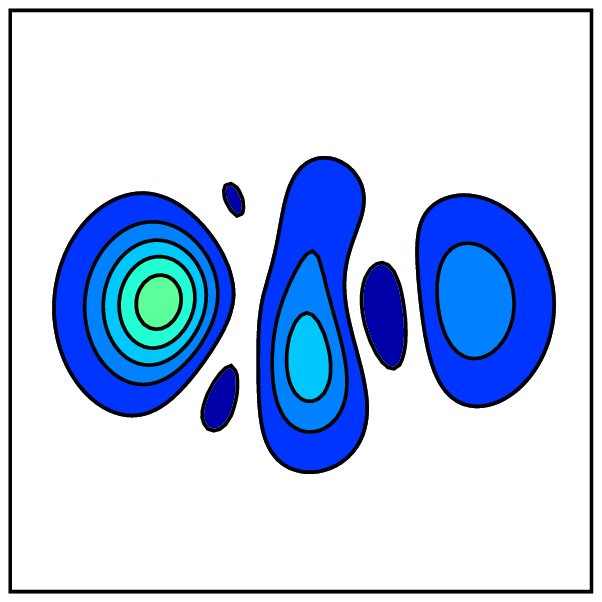} & \\
	\begin{minipage}{0.9cm}\vspace{-1.7cm}(d) \end{minipage} &
	\includegraphics[width=1.8cm, bb=223 311 394 481, clip]{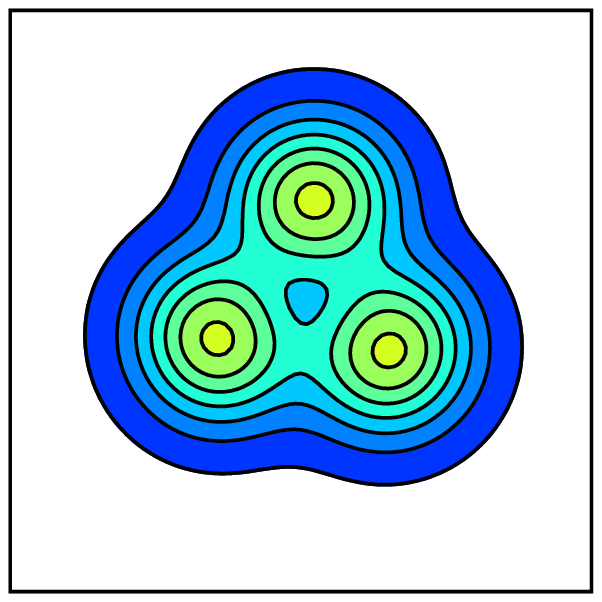} &
	\includegraphics[width=1.8cm, bb=223 311 394 481, clip]{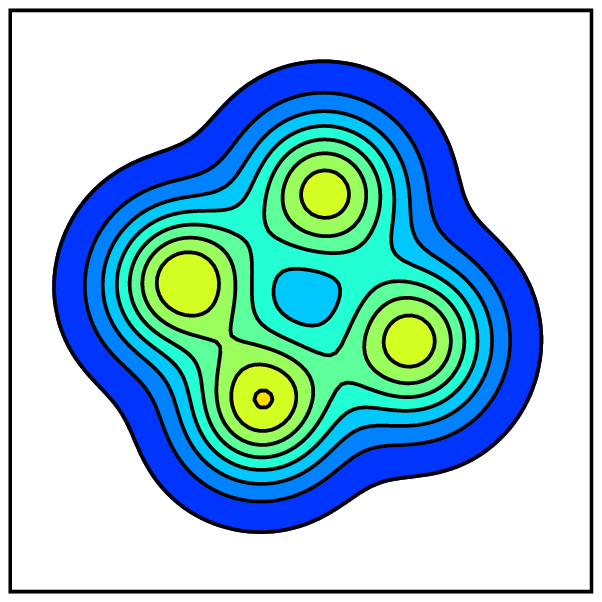} &
	\includegraphics[width=1.8cm, bb=223 311 394 481, clip]{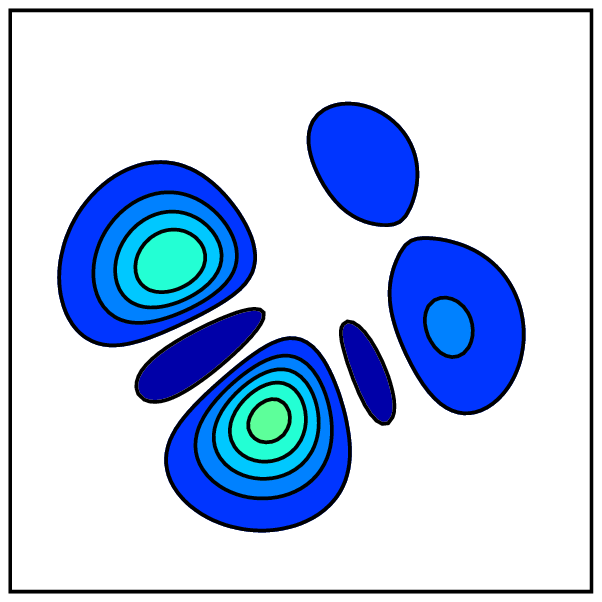} \\
	\begin{minipage}{0.9cm}\vspace{-1.7cm}(e) \end{minipage} &
	\includegraphics[width=1.8cm, bb=223 311 394 481, clip]{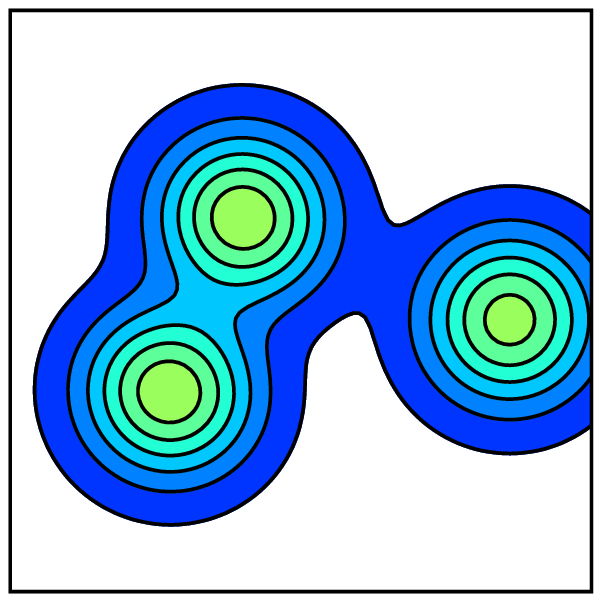} &
	\includegraphics[width=1.8cm, bb=223 311 394 481, clip]{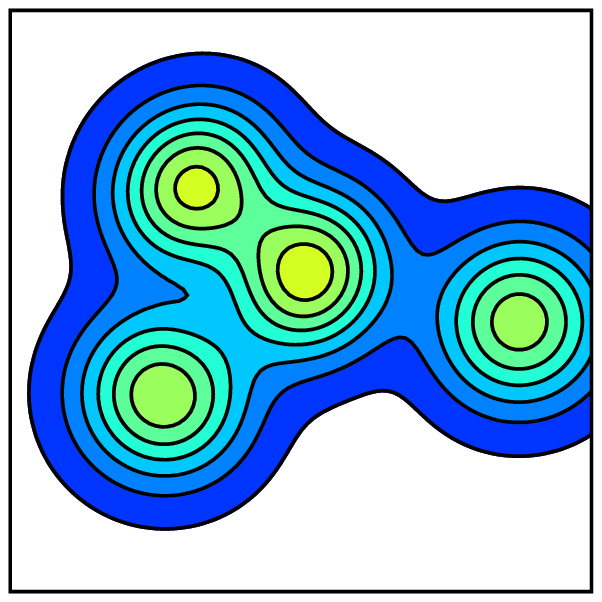} &
	\includegraphics[width=1.8cm, bb=223 311 394 481, clip]{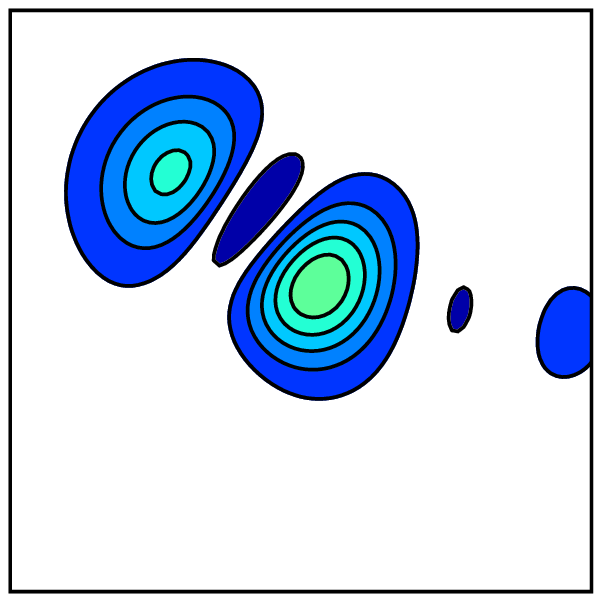} \\
	\begin{minipage}{0.9cm}\vspace{-1.7cm}(f) \end{minipage} &
	\includegraphics[width=1.8cm, bb=223 311 394 481, clip]{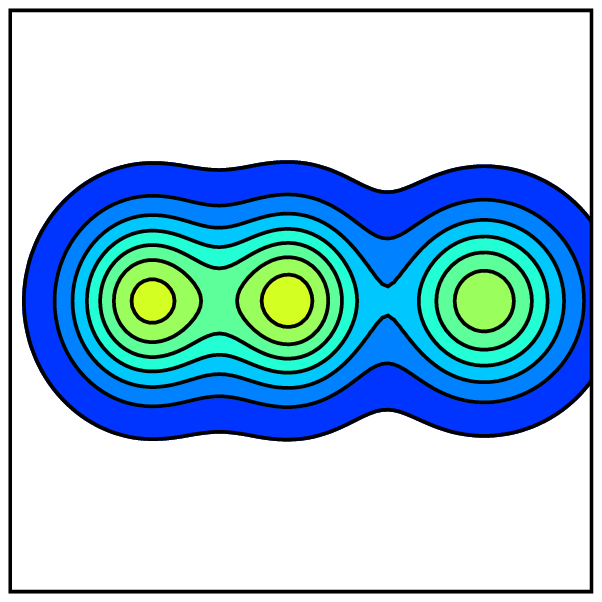} &
	\includegraphics[width=1.8cm, bb=223 311 394 481, clip]{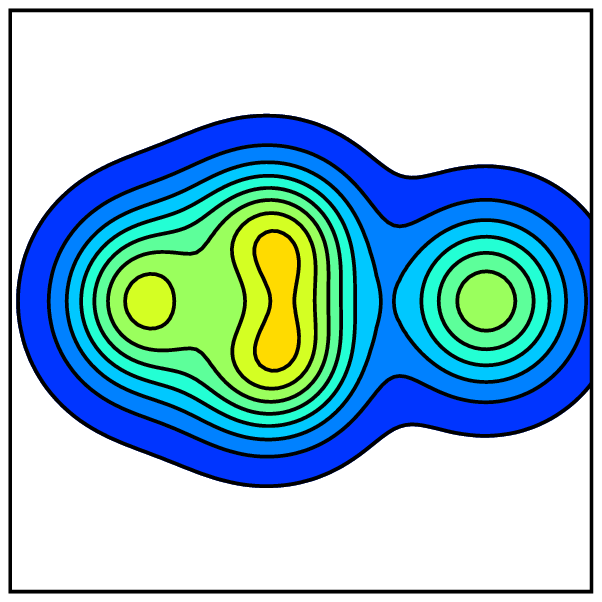} &
	\includegraphics[width=1.8cm, bb=223 311 394 481, clip]{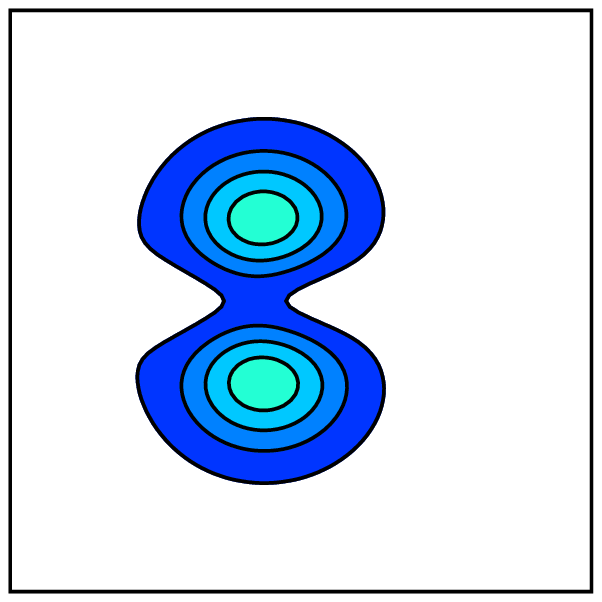} \\
	\end{tabular}}
	\caption{(Color online) Density distributions of the intrinsic wave functions for the positive-parity states of $^{14}$C.
	The proton density $\Tilde{\rho}_{p}$, neutron density $\Tilde{\rho}_{n}$, 
	and difference between the neutron and proton densities 
	$\Tilde{\rho}_{n} - \Tilde{\rho}_{p}$ are illustrated in the left, middle, and right columns, respectively.
	The density distributions of the intrinsic wave functions at 
	(a) $(\beta \cos \gamma,\beta \sin \gamma)=(0.00,0.00)$,
	(b) $(\beta \cos \gamma,\beta \sin \gamma)=(0.23,0.04)$,
	(c) $(\beta \cos \gamma,\beta \sin \gamma)=(0.45,0.17)$,
	(d) $(\beta \cos \gamma,\beta \sin \gamma)=(0.25,0.35)$,
	(e) $(\beta \cos \gamma,\beta \sin \gamma)=(0.78,0.22)$, and 
	(f) $(\beta \cos \gamma,\beta \sin \gamma)=(0.93,0.04)$ on the $\beta$-$\gamma$ plane are shown. 
	The size of the box is 10 $\times$ 10 fm$^{2}$.}
	\label{density_14C_+}
\end{figure}

First, we discuss the intrinsic structures of positive-parity states of $^{14}$C.
The density distributions of the intrinsic wave functions for positive-parity states 
are illustrated in Fig.~\ref{density_14C_+}.
The energy minimum state at $(\beta \cos \gamma, \beta \sin \gamma)=(0.00, 0.00)$ 
in the positive-parity energy surface shows almost spherical density distributions 
as seen in Fig.~\ref{density_14C_+}(a).
In this wave function, the centers of the single-particle Gaussian wave packets gather near the origin.
That is, this state has no spatially developed cluster structure 
and it is almost equivalent to the shell-model state 
with the $p_{3/2}$-subshell closed-proton configuration and the $p$-shell closed-neutron configuration.
The density distribution for the minimum point 
$(\beta \cos \gamma, \beta \sin \gamma)=(0.23, 0.04)$ 
in the $0^{+}$ energy surface is shown in Fig.~\ref{density_14C_+}(b).
This wave function is found to be the dominant component of the ground state obtained by the GCM calculation
as is shown later.
In this state, an $\alpha$ cluster core somewhat develops compared with the state (a) 
for the energy minimum before the total-angular-momentum projection.
However, single-particle Gaussian wave packets still gather around the origin; therefore, 
this state is regarded as the intermediate between the shell-model structure and the cluster structure.
This feature is also reflected in the expectation value of squared intrinsic spin of protons $\langle \hat{S}_{p}^{2}\rangle = 0.77$, 
which is smaller than the ideal value $4/3$ for the $p_{3/2}$-subshell closed configuration and 
larger than the value $0$ for a 3$\alpha$-cluster configuration.

Figure~\ref{density_14C_+}(c) indicates the density distribution for a triaxial deformed state.
In this state, to be clear from the proton density 
and also from the expectation values of the squared intrinsic proton spin $\langle \hat{S}_{p}^{2} \rangle = 0.15$,
three $\alpha$-cluster cores develop.
Excess neutrons occupy the $sd$-like orbitals
distributing along the bisector of the vertical angle of the isosceles-triangle which consists of three $\alpha$ clusters.
These neutron orbitals make this state triaxial.
After the GCM calculation, a $K^{\pi}=0^{+}$ band and its $K^{\pi}=2^{+}$ side band dominated by this triaxial state are obtained.

In the large deformation region, three $\alpha$-cluster cores develop well in $^{14}$C. 
Various configurations of three $\alpha$ clusters appear, depending on the deformation parameters, $\beta$ and $\gamma$, 
as seen in Figs.~\ref{density_14C_+}(d), (e), and (f) for typical density distributions 
of oblate, triaxial, and prolate deformed states, respectively.
It is found that the equilateral-triangular structure, obtuse-angle-triangular structure,
and linear-chain structure of three $\alpha$ clusters arise
in the oblate state (d), the triaxial state (e), and the prolate state (f), respectively.
In these states, the expectation values of squared intrinsic proton spin $\langle \hat{S}_{p}^{2}\rangle$ 
are smaller than 0.10, which is almost consistent with the value for three $(0s)^{4}$ $\alpha$ cores.
It is interesting that the excess neutrons distribute 
around two of the three $\alpha$ clusters in the triaxial state (e) and prolate state (f), 
which indicates $^{10}$Be+$\alpha$ correlations. 
In the viewpoint of two-center $^{10}$Be+$\alpha$ cluster structure,
the degree of the freedom for rotation of the deformed $^{10}$Be cluster is described on the $\beta$-$\gamma$ plane.
We mention here that the linear-chain structure (f) is the base 
of the flat energy region at the large prolate deformation.
After the GCM calculation, the linear-chain state (f) constructs a rotational band.

\begin{figure}[tb]
	\centering
	{\tabcolsep=0.7mm
	\begin{tabular}{ccccc}
	\vspace{0.3cm}
	 &
	\hspace{-0.2cm} $\Tilde{\rho}_{p}$ & \hspace{-0.2cm} $\Tilde{\rho}_{n}$ 
	& \hspace{-0.15cm}  $\Tilde{\rho}_{n} - \Tilde{\rho}_{p}$ & 
	\multirow{4}{*}{\begin{minipage}{1.8cm}[1/fm$^{2}$] \includegraphics[width=1.8cm, bb=379 301 450 490, clip]{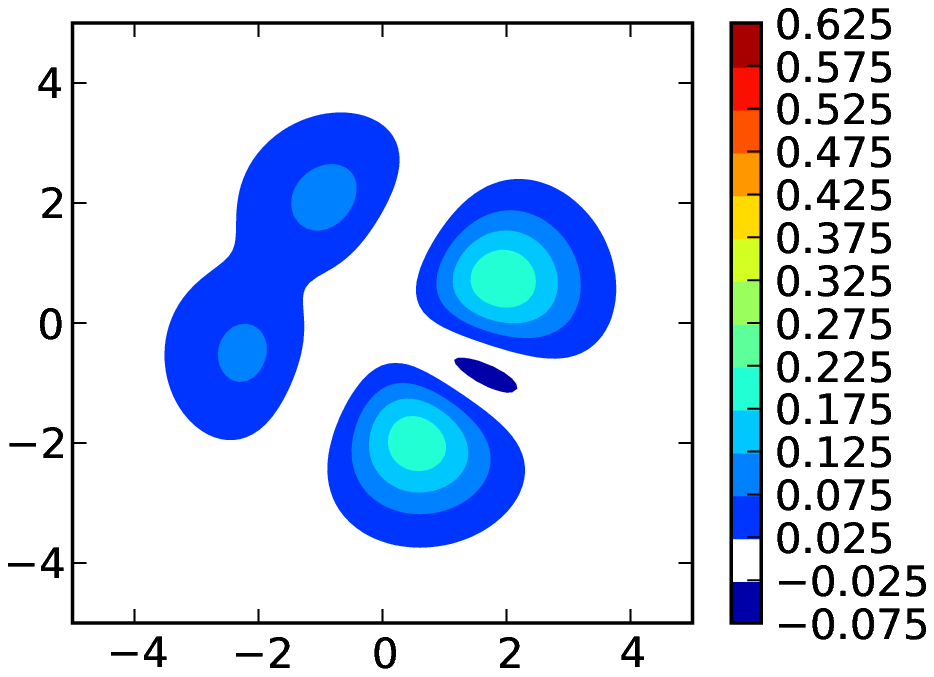}\end{minipage}} \\
	\begin{minipage}{0.9cm}\vspace{-1.7cm}(a) \end{minipage} &
	\includegraphics[width=1.8cm, bb=223 311 394 481, clip]{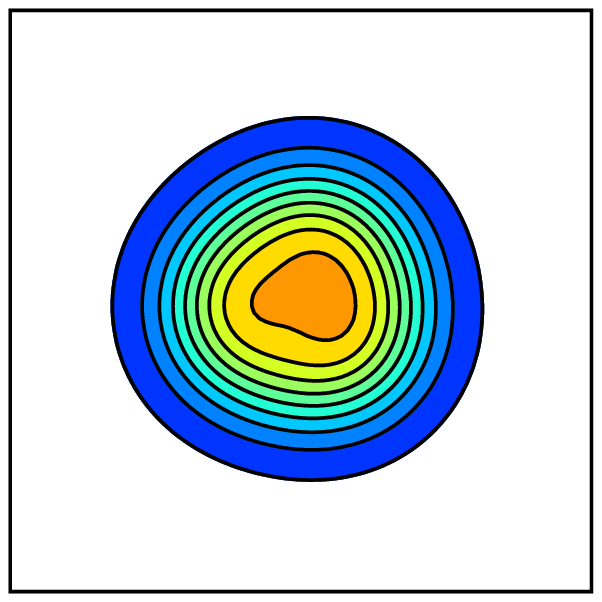} &
	\includegraphics[width=1.8cm, bb=223 311 394 481, clip]{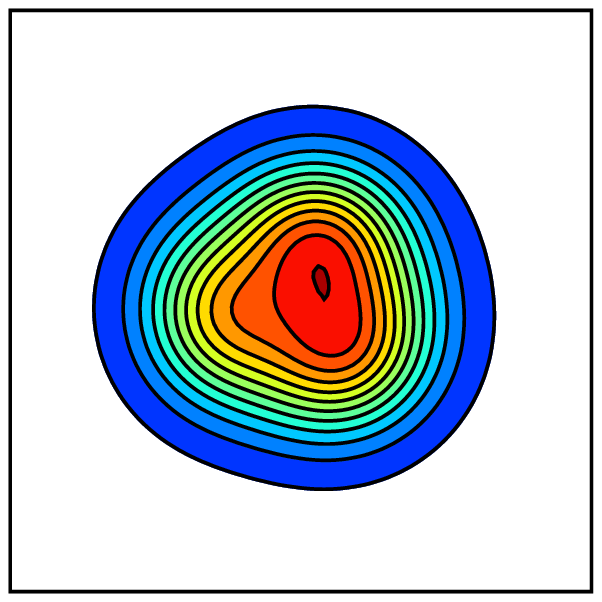} &
	\includegraphics[width=1.8cm, bb=223 311 394 481, clip]{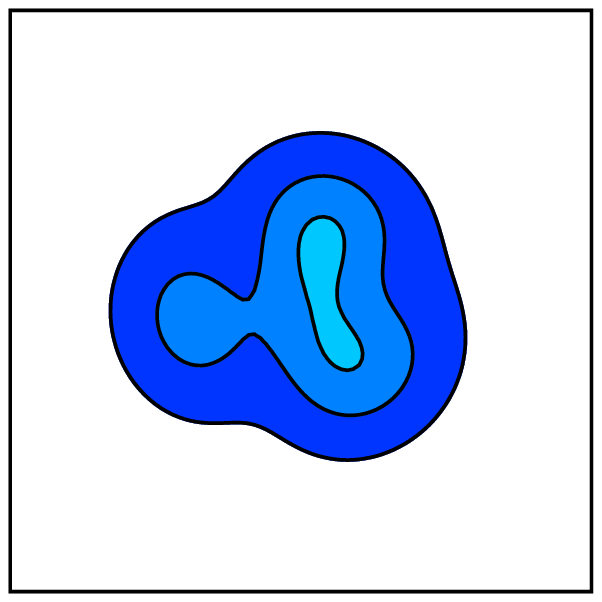} & \\
	\begin{minipage}{0.9cm}\vspace{-1.7cm}(b) \end{minipage} &
	\includegraphics[width=1.8cm, bb=223 311 394 481, clip]{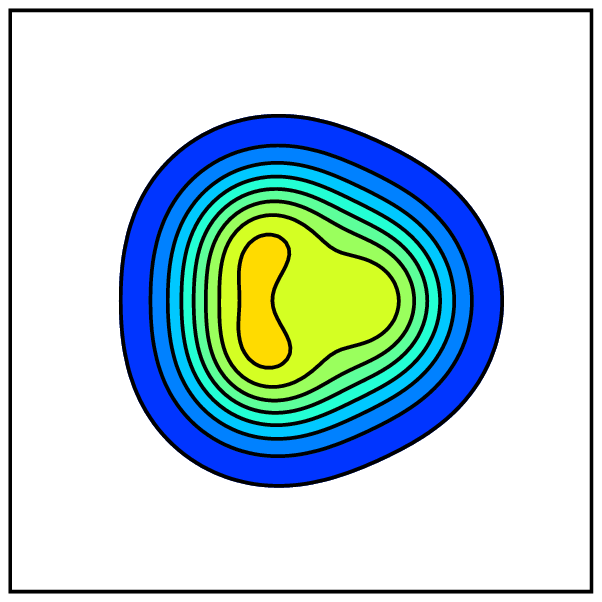} &
	\includegraphics[width=1.8cm, bb=223 311 394 481, clip]{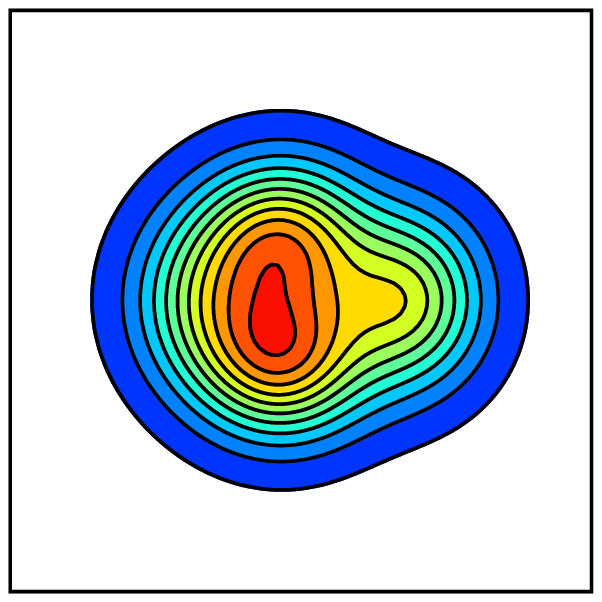} &
	\includegraphics[width=1.8cm, bb=223 311 394 481, clip]{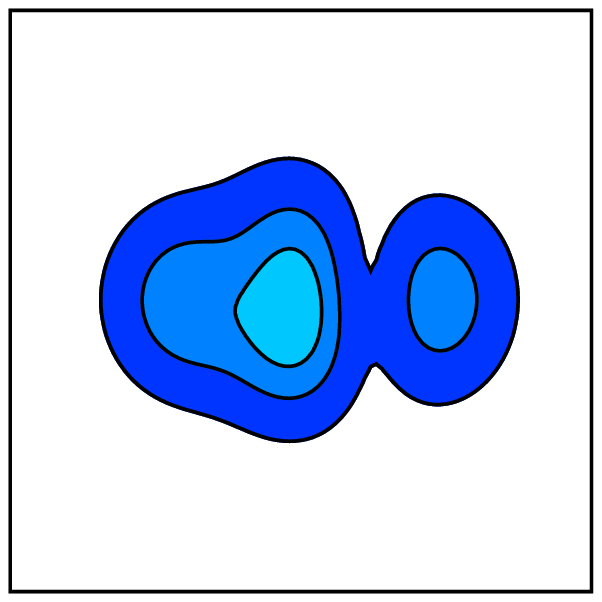} & \\
	\begin{minipage}{0.9cm}\vspace{-1.7cm}(c) \end{minipage} &
	\includegraphics[width=1.8cm, bb=223 311 394 481, clip]{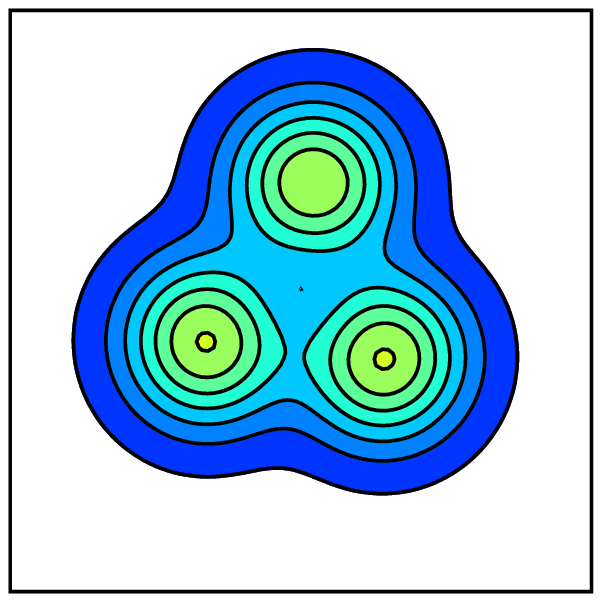} &
	\includegraphics[width=1.8cm, bb=223 311 394 481, clip]{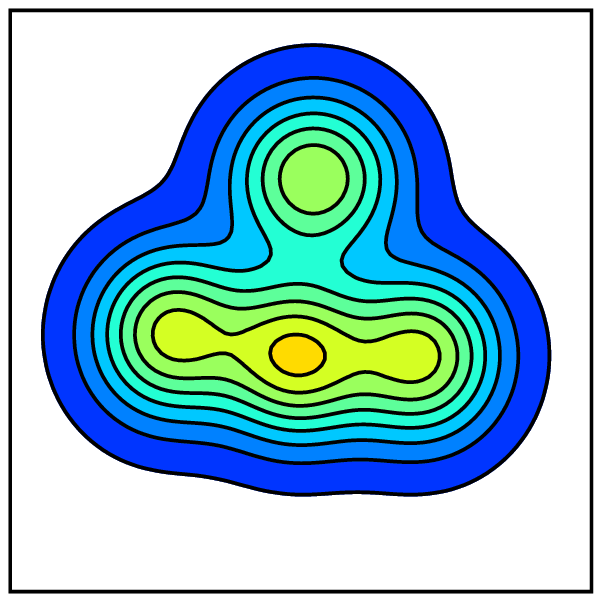} &
	\includegraphics[width=1.8cm, bb=223 311 394 481, clip]{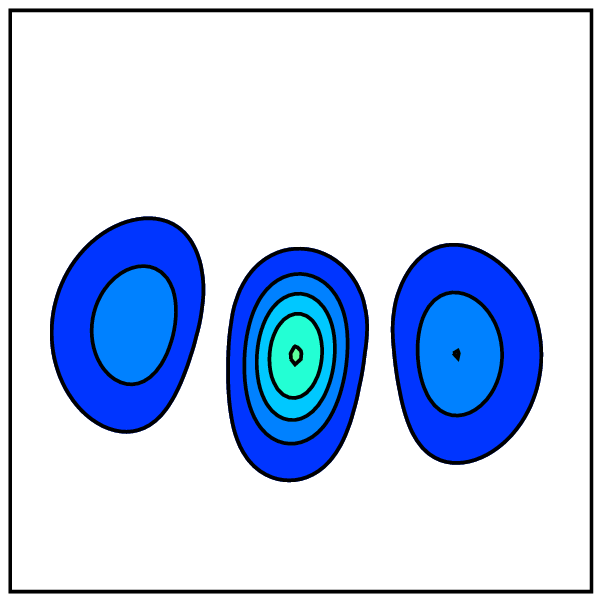} \\
	\begin{minipage}{0.9cm}\vspace{-1.7cm}(d) \end{minipage} &
	\includegraphics[width=1.8cm, bb=223 311 394 481, clip]{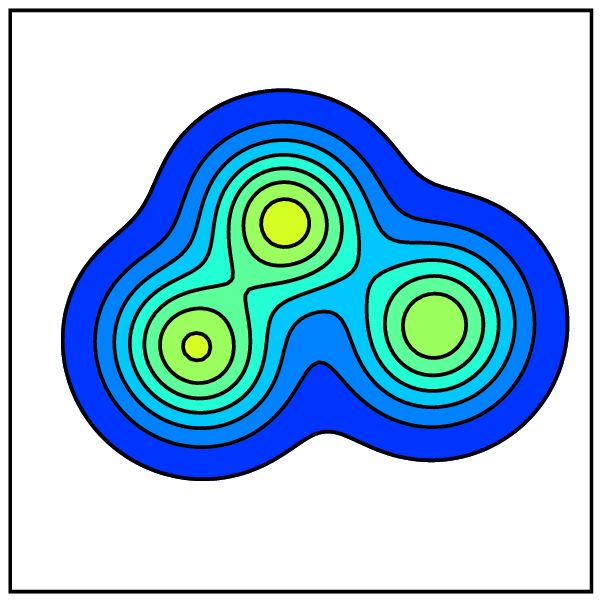} &
	\includegraphics[width=1.8cm, bb=223 311 394 481, clip]{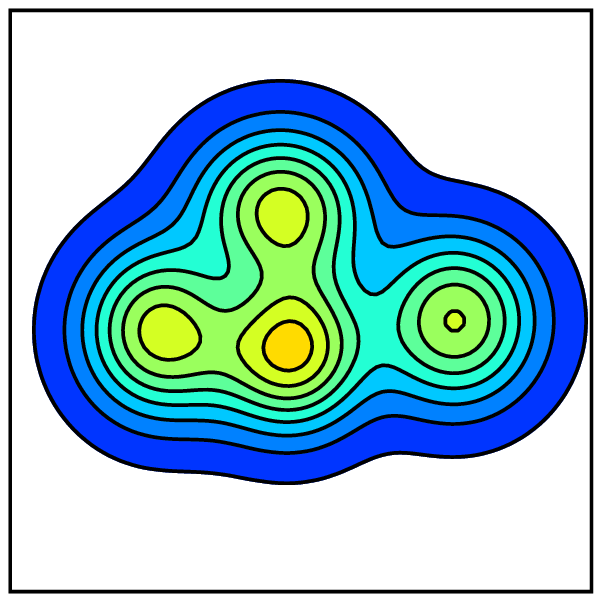} &
	\includegraphics[width=1.8cm, bb=223 311 394 481, clip]{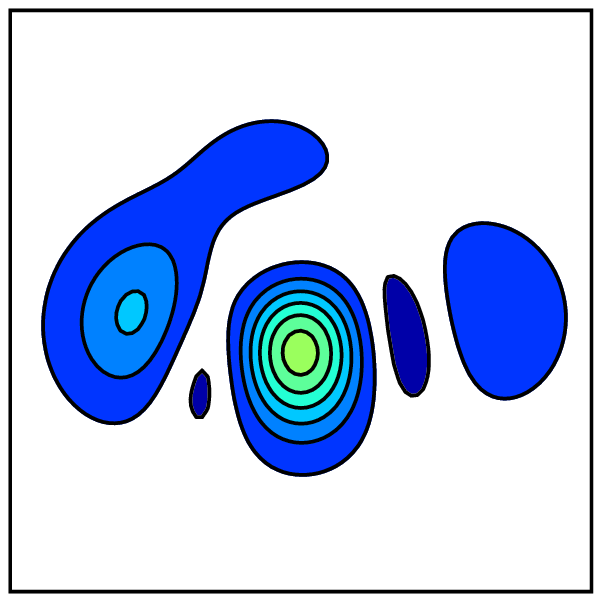} \\
	\begin{minipage}{0.9cm}\vspace{-1.7cm}(e) \end{minipage} &
	\includegraphics[width=1.8cm, bb=223 311 394 481, clip]{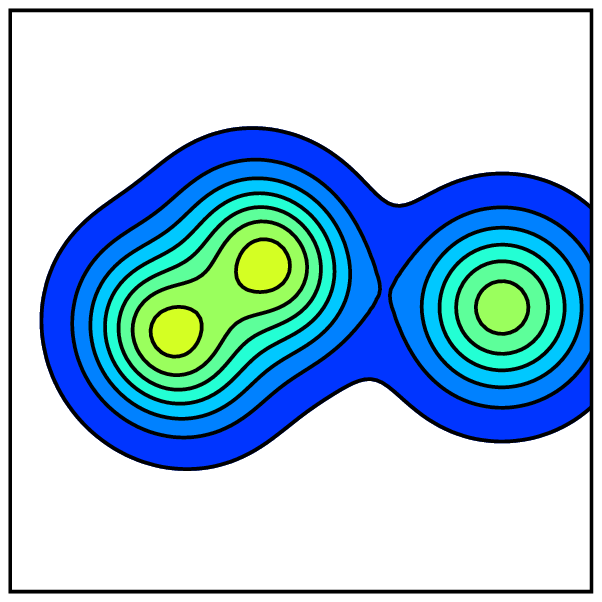} &
	\includegraphics[width=1.8cm, bb=223 311 394 481, clip]{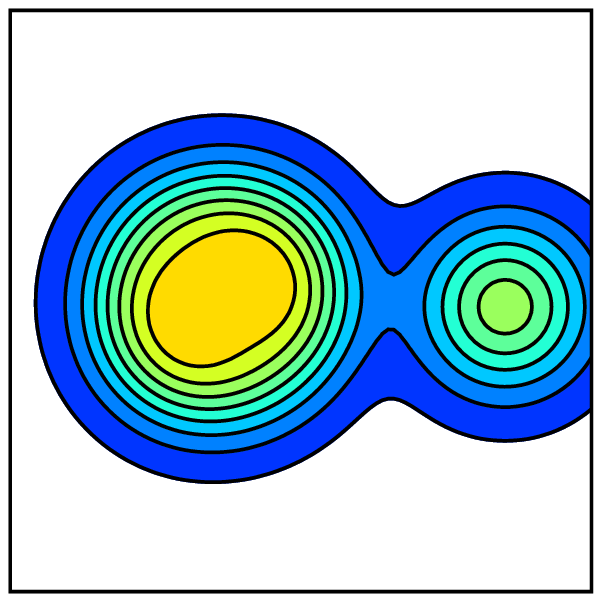} &
	\includegraphics[width=1.8cm, bb=223 311 394 481, clip]{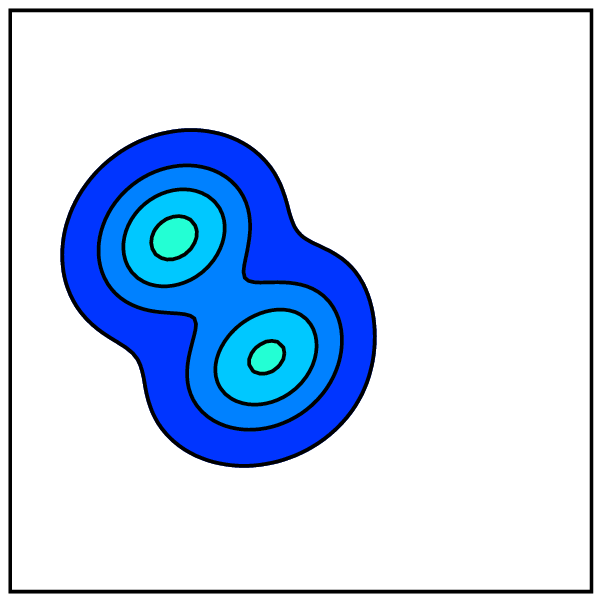} \\
	\end{tabular}}
	\caption{(Color online) Density distributions of the intrinsic wave functions for the negative-parity states of $^{14}$C.
	The proton density $\Tilde{\rho}_{p}$, neutron density $\Tilde{\rho}_{n}$, 
	and difference between the neutron and proton densities 
	$\Tilde{\rho}_{n} - \Tilde{\rho}_{p}$ are illustrated in the left, middle, and right columns, respectively.
	The density distributions of the intrinsic wave functions at 
	(a) $(\beta \cos \gamma,\beta \sin \gamma)=(0.08,0.04)$,
	(b) $(\beta \cos \gamma,\beta \sin \gamma)=(0.20,0.09)$,
	(c) $(\beta \cos \gamma,\beta \sin \gamma)=(0.25,0.35)$,
	(d) $(\beta \cos \gamma,\beta \sin \gamma)=(0.60,0.17)$, and 
	(e) $(\beta \cos \gamma,\beta \sin \gamma)=(0.93,0.04)$ on the $\beta$-$\gamma$ plane are shown. 
	The size of the box is 10 $\times$ 10 fm$^{2}$.}
	\label{density_14C_-}
\end{figure}

Next, we discuss the intrinsic structure of negative-parity states. 
The density distributions for negative-parity states are illustrated in Fig.~\ref{density_14C_-}.
The density distribution of the intrinsic wave function at the minimum point $(\beta \cos \gamma, \beta \sin \gamma)=(0.08, 0.04)$ 
in the negative-parity energy surface is shown Fig.~\ref{density_14C_-}(a).
In this wave function, the centers of the single-particle Gaussian wave packets 
gather near the origin and there is no spatially developed cluster structure.
In this state, the expectation value of squared intrinsic spin of protons is 1.24, which is close to the ideal value $4/3$ for 
the $p_{3/2}$-subshell closed configuration, 
while that of neutrons is 0.85, which deviates from the $p$-shell closed configuration value $0$.
Therefore, this state is interpreted as the shell-model structure having 
$1p$-$1h$ excitations on the neutron $p$ shell with the almost $p_{3/2}$-subshell closed-proton configuration.
Compared with an almost spherical shape of the state (a), 
an $\alpha$ cluster core begins to develop slightly at the minimum point 
$(\beta \cos \gamma, \beta \sin \gamma)=(0.20, 0.09)$ in the $3^{-}$ energy surface (Fig.~\ref{density_14C_-}(b)).
However, single-particle Gaussian wave packets still gather around the origin, and this state
is regarded as the intermediate between the shell-model structure and the cluster structure.

Figures~\ref{density_14C_-}(c), (d), and (e) show intrinsic density distributions
for typical deformed states---oblate, triaxial, and prolate, respectively.
It is found that the isosceles-triangular structure and obtuse-angle-triangular structure 
of three $\alpha$-cluster cores arise in the oblate state (c) and triaxial state (d), respectively, as shown in the proton densities.
These structures are similar to those obtained for the positive-parity states.

However, the structure of the prolate deformed states is somewhat different from the positive-parity states.
In the prolate state (e), the proton density $\Tilde{\rho}_{p}$ has three peaks, which indicate three $\alpha$-cluster cores.
However, the expectation values of squared intrinsic proton spin $\langle \hat{S}_{p}^{2} \rangle = 0.33$ 
indicates that the prolate state (e) contains components of the $(0s)^4$ $\alpha$ breaking.
Moreover, three $\alpha$ clusters shows a bending structure
in stead of the straight-line 3$\alpha$ configuration seen 
in the linear-chain structure (Fig.~\ref{density_14C_+}(f)) of the positive-parity prolate state.
That is, the $^{10}$Be cluster rotates considerably
though the prolate state (e) has almost the axial symmetric shape 
caused by the constraint parameters close to the $\gamma = 0^{\circ}$ axis on the $\beta$-$\gamma$ plane.
This indicates that the linear-chain structure is less favored 
in the negative-parity states than in the positive-parity states.
It is consistent with the features of the $\beta$-$\gamma$ energy surfaces, 
where there is no plateau in the negative-parity states as mentioned before.

Thus, on the $\beta$-$\gamma$ plane for the $^{14}$C system, 
we obtained various structures such as 
the shell-model structure in the small $\beta$ region
and the cluster structure in the large $\beta$ region.
In particular, various spatial configurations of three $\alpha$ clusters such as 
the linear-chain and the equilateral-triangular structures are obtained 
as a function of the triaxiality $\gamma$.
These basis wave functions are proper
for the study of $^{14}$C where cluster and shell-model aspects are expected to coexist.

\subsection{Energy levels}

To calculate the energy spectra of $^{14}$C, 
we superposed the wave functions obtained by the $\beta$-$\gamma$ constraint AMD 
at 196 mesh points on the $\beta$-$\gamma$ plane by using the GCM.

\begin{figure}[tb]
\centering
	\includegraphics[angle=-90, width=8.6cm, bb=290 54 542 392, clip]{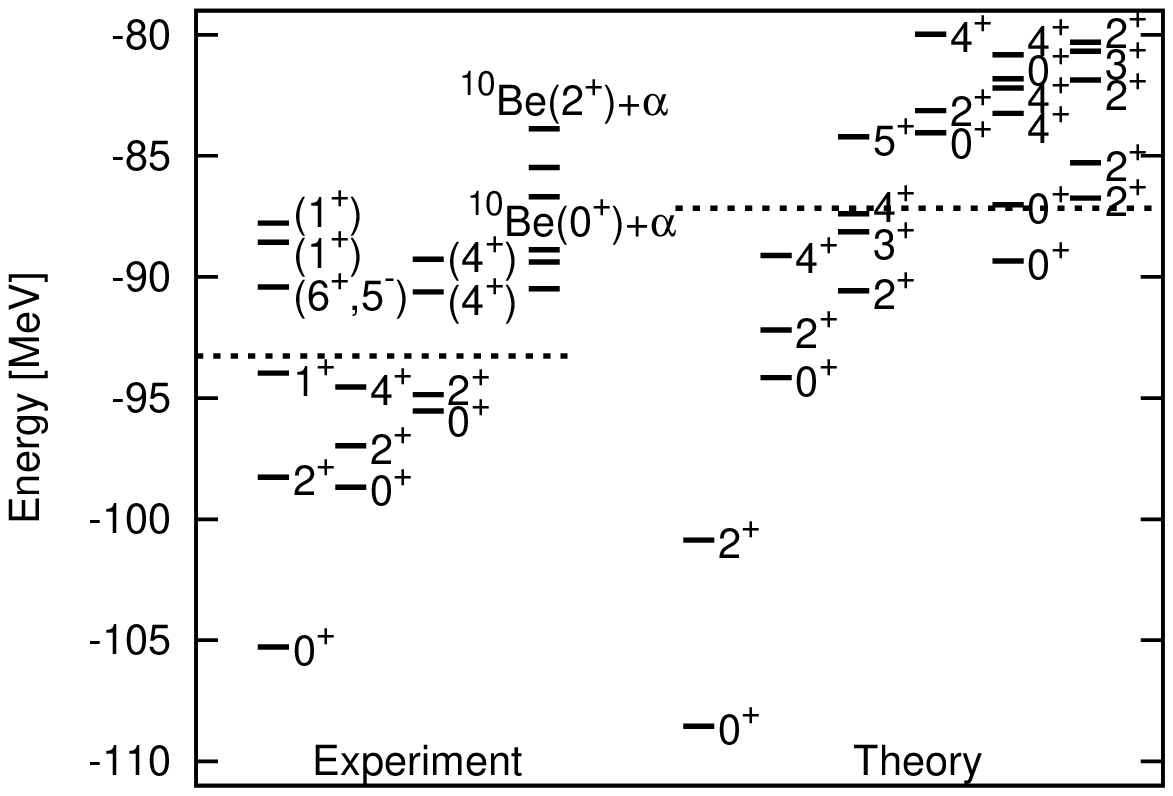}
	\caption{Energy levels of the positive-parity states in $^{14}$C.
		Four columns on the left are the experimental data and six columns on the right are the calculated results.
		The dotted lines on the left and right show 
		the experimental and theoretical $^{10}$Be$+ \alpha$ threshold energies, respectively.}
\label{Energy_level_14C_+}
\end{figure}

\begin{table}[tb]
	\caption{Electromagnetic transition strengths $B(E2)$ which are stronger than 1.0 $e^{2}$fm$^{4}$ for the positive-parity states in $^{14}$C.
	The unit is $e^{2}$fm$^{4}$.}
	\label{E2_14C_+}
	\begin{tabular}[t]{ccc} \hline \hline
\multirow{2}{*}{Transition} & \multicolumn{2}{c}{Strength} \\ \cline{2-3}
 & Theory & Experiment \\ \hline
$2^{+}_{1} \rightarrow 0^{+}_{1}$ &  4.4 & $3.6 \pm 0.6$ \\
$2^{+}_{1} \rightarrow 0^{+}_{3}$ &  1.4 & \\
$2^{+}_{2} \rightarrow 0^{+}_{2}$ &  6.0 & \\
$2^{+}_{3} \rightarrow 0^{+}_{2}$ &  1.0 & \\
$2^{+}_{4} \rightarrow 0^{+}_{4}$ &  4.1 & \\
$2^{+}_{5} \rightarrow 0^{+}_{3}$ &  1.3 & \\
$2^{+}_{6} \rightarrow 0^{+}_{2}$ &  1.3 & \\
$2^{+}_{6} \rightarrow 0^{+}_{4}$ &  3.8 & \\
$2^{+}_{6} \rightarrow 0^{+}_{5}$ & 67.2 & \\
$2^{+}_{6} \rightarrow 0^{+}_{6}$ &  5.3 & \\
$2^{+}_{7} \rightarrow 0^{+}_{4}$ &  1.6 & \\
$2^{+}_{7} \rightarrow 0^{+}_{5}$ &  1.8 & \\
$2^{+}_{7} \rightarrow 0^{+}_{6}$ &  2.8 & \\
$2^{+}_{8} \rightarrow 0^{+}_{6}$ & 14.3 & \\
 & & \\
$2^{+}_{3} \rightarrow 2^{+}_{2}$ &  7.3 & \\
$2^{+}_{5} \rightarrow 2^{+}_{1}$ &  3.3 & \\
$2^{+}_{6} \rightarrow 2^{+}_{2}$ &  1.7 & \\
$2^{+}_{6} \rightarrow 2^{+}_{4}$ &  2.6 & \\
$2^{+}_{6} \rightarrow 2^{+}_{5}$ &  1.3 & \\
$2^{+}_{7} \rightarrow 2^{+}_{6}$ &  6.3 & \\
$2^{+}_{8} \rightarrow 2^{+}_{3}$ &  1.1 & \\
$2^{+}_{8} \rightarrow 2^{+}_{7}$ &  2.4 & \\
 & & \\
$3^{+}_{1} \rightarrow 2^{+}_{2}$ &  2.5 & \\
$3^{+}_{1} \rightarrow 2^{+}_{3}$ & 10.9 & \\
$3^{+}_{1} \rightarrow 2^{+}_{4}$ &  1.7 & \\
$3^{+}_{2} \rightarrow 2^{+}_{7}$ &  3.2 & \\
$3^{+}_{2} \rightarrow 2^{+}_{8}$ &  2.0 & \\
	\hline \hline
	\end{tabular} \qquad
	\begin{tabular}[t]{ccc} \hline \hline
\multirow{2}{*}{Transition} & \multicolumn{2}{c}{Strength} \\ \cline{2-3}
 & Theory & Experiment \\ \hline
$4^{+}_{1} \rightarrow 2^{+}_{2}$ &  4.7 & \\
$4^{+}_{2} \rightarrow 2^{+}_{2}$ &  2.8 & \\
$4^{+}_{2} \rightarrow 2^{+}_{3}$ &  2.1 & \\
$4^{+}_{3} \rightarrow 2^{+}_{3}$ &  1.0 & \\
$4^{+}_{3} \rightarrow 2^{+}_{4}$ &  5.1 & \\
$4^{+}_{3} \rightarrow 2^{+}_{6}$ &  2.4 & \\
$4^{+}_{4} \rightarrow 2^{+}_{6}$ & 12.9 & \\
$4^{+}_{4} \rightarrow 2^{+}_{8}$ &  1.4 & \\
$4^{+}_{5} \rightarrow 2^{+}_{2}$ &  2.5 & \\
$4^{+}_{5} \rightarrow 2^{+}_{4}$ &  1.4 & \\
$4^{+}_{5} \rightarrow 2^{+}_{6}$ &  4.4 & \\
$4^{+}_{5} \rightarrow 2^{+}_{8}$ &  5.8 & \\
$4^{+}_{6} \rightarrow 2^{+}_{4}$ &  1.0 & \\
$4^{+}_{6} \rightarrow 2^{+}_{6}$ & 88.3 & \\
$4^{+}_{6} \rightarrow 2^{+}_{7}$ &  8.7 & \\
 & & \\
$4^{+}_{1} \rightarrow 3^{+}_{1}$ &  1.2 & \\
$4^{+}_{2} \rightarrow 3^{+}_{1}$ &  4.6 & \\
$4^{+}_{3} \rightarrow 3^{+}_{1}$ &  3.5 & \\
$4^{+}_{4} \rightarrow 3^{+}_{2}$ &  2.8 & \\
 & & \\
$4^{+}_{2} \rightarrow 4^{+}_{1}$ &  4.5 & \\
$4^{+}_{3} \rightarrow 4^{+}_{2}$ &  1.7 & \\
$4^{+}_{4} \rightarrow 4^{+}_{3}$ &  1.5 & \\
$4^{+}_{6} \rightarrow 4^{+}_{3}$ &  3.0 & \\
$4^{+}_{6} \rightarrow 4^{+}_{4}$ &  6.0 & \\
$4^{+}_{6} \rightarrow 4^{+}_{5}$ &  2.1 & \\
 & & \\
$5^{+}_{1} \rightarrow 3^{+}_{1}$ &  2.4 & \\
 & & \\
$5^{+}_{1} \rightarrow 4^{+}_{1}$ &  4.9 & \\
$5^{+}_{1} \rightarrow 4^{+}_{2}$ &  4.5 & \\
	\hline \hline
	\end{tabular}
\end{table}

\begin{table}[tb]
	\caption{Root-mean-square radii for mass distributions of the positive-parity states in $^{14}$C. The unit is fm.}
	\label{Radius_14C_+}
	\centering
	\begin{tabular}{ccc} \hline \hline
\multirow{2}{*}{State} & \multicolumn{2}{c}{Radius} \\ \cline{2-3}
 & Theory & Experiment \\ \hline
	$0^{+}_{1}$ & 2.22 & $2.30 \pm 0.07$ \\
	$0^{+}_{2}$ & 2.45 & \\
	$0^{+}_{3}$ & 2.31 & \\ 
	$0^{+}_{4}$ & 2.46 & \\ 
	$0^{+}_{5}$ & 2.82 & \\ 
	$0^{+}_{6}$ & 2.69 & \\ 
	\hline \hline
	\end{tabular}
\end{table}

First, we describe the results for the positive-parity states.
We show the calculated positive-parity energy levels in Fig.~\ref{Energy_level_14C_+}
as well as the experimental levels.
In the three columns from the left, we display the experimental energy levels for all the positive-parity assigned states \cite{Ajzenberg_A=13-15_91}.
In the fourth column, we show the states observed recently
in $^{10}$Be+$\alpha$ breakup reactions \cite{Milin_14C_04, Soic_CandBisotopes_04, Price_14C_07, Haigh_14C_08}.
The lower three of them are the states with the strong population of 
the $^{10}$Be$(0^{+}_{1})$+$\alpha$ and the upper three are those of 
the $^{10}$Be$(2^{+}_{1})$+$\alpha$.
In the theoretical results, we classified the calculated states in five groups by analyzing 
the components of the basis wave functions and $E2$ transition strengths.
The calculated $E2$ transition strengths and root-mean-square radii 
are listed in Tables~\ref{E2_14C_+} and \ref{Radius_14C_+}, respectively, along with experimental data \cite{Ajzenberg_A=13-15_91,Ozawa_rmsRadius_01}.

The calculated $0^{+}_{1}$ and $2^{+}_{1}$ states constitute the ground band. 
The calculations reproduce well the experimental data for the properties of the ground band 
such as $B(E2, 2^{+}_{1} \rightarrow 0^{+}_{1})$ and the root-mean-square radius of the $0^{+}_{1}$ state.
The results show that 
the basis wave function at the minimum point $(\beta \cos \gamma, \beta \sin \gamma) = (0.23, 0.04)$ in the $0^{+}$ energy surface
dominates these ground-band states.
For instance, the overlap with the $0^{+}_{1}$ state is 97\%.
As discussed before, this dominant basis wave function has 
no spatial development of three $\alpha$ clusters but the intermediate structure (Fig.~\ref{density_14C_+}(b)).

The calculated $0^{+}_{2}$, $2^{+}_{2}$, and $4^{+}_{1}$ states in the $K^{\pi} = 0^{+}_{2}$ band 
have large overlap with the AMD base at $(\beta \cos \gamma,\beta \sin \gamma) = (0.45,0.17)$ as
78\% overlap in the $0^{+}_{2}$ state.
The $2^{+}_{3}$ ,$3^{+}_{1}$, $4^{+}_{2}$, and $5^{+}_{1}$ states also contain dominant components of the same basis wave function.
As mentioned before, this basis wave function has the triaxial intrinsic structure 
with three $\alpha$ clusters and excess neutrons occupying the $sd$-like longitudinal orbitals (Fig.~\ref{density_14C_+}(c)).
Therefore, it is natural to interpret these states as the 
$K^{\pi}=0^{+}$ band and the $K^{\pi}=2^{+}$ side band which are constructed from the triaxially deformed intrinsic state.
It is an interesting suggestion that the side band is built on the $0^{+}_{2}$ band of $^{14}$C 
owing to the triaxial shape caused by the excess neutron distributions around the 3$\alpha$ core.
The calculated $E2$ strengths for the intraband transitions are comparable to that in the ground band.
Because there is some mixing of $K$ components in these bands,
$E2$ transitions from $4^{+}_{2}$ and $5^{+}_{1}$ states are scattered.
Although the excitation energy of the band-head $0^{+}_{2}$ state is much higher than the experimental energy of the $0^{+}_{2}$ state, 
the calculated $0^{+}_{2}$, $2^{+}_{2}$, and $4^{+}_{1}$ states may correspond to
the experimental $0^{+}_{2}$, $2^{+}_{2}$, and $4^{+}_{1}$ states when we measure the energies from the $^{10}$Be+$\alpha$ threshold.
To conclude the correspondence with the experimental levels, 
we need more detailed experimental data such as transition strengths.

For the calculated $0^{+}_{5}$, $2^{+}_{6}$, and $4^{+}_{6}$ states,
we found that the main component comes from the AMD wave function
at $(\beta \cos \gamma,\beta \sin \gamma)=(0.93,0.04)$ 
as 64\% overlap in the $0^{+}_{5}$ state.
As discussed before, this dominant wave function has the linear-chain structure of three $\alpha$-cluster cores 
with the $^{10}$Be+$\alpha$ correlation (Fig.~\ref{density_14C_+}(f)).
In the viewpoint of two-center $^{10}$Be+$\alpha$ cluster structure,
the rotational mode of the deformed $^{10}$Be cluster is at least partially incorporated 
in the GCM on the $\beta$-$\gamma$ plane. 
For example, the basis wave function at $(\beta \cos \gamma,\beta \sin \gamma)=(0.78,0.22)$ in the finite $\gamma$ region shows a 
tilting $^{10}$Be cluster configuration (Fig.~\ref{density_14C_+}(e)).
In the results of the GCM calculations, the $0^{+}_{5}$ has only 9\% overlap 
with this tilting $^{10}$Be configuration, which is much smaller than 
the overlap with the linear-chain structure at $(\beta \cos \gamma,\beta \sin \gamma)=(0.93,0.04)$. 
In other words, the GCM amplitudes concentrate on the straight-line $^{10}$Be configuration rather than the tilting one.
As a result of the large deformation of the linear-chain structure of three $\alpha$ clusters, 
the $E2$ transitions in this band are strong as
$B(E2, 2^{+}_{6} \rightarrow 0^{+}_{5})=67.2$ $e^{2}$fm$^{4}$ and
$B(E2, 4^{+}_{6} \rightarrow 2^{+}_{6})=88.3$ $e^{2}$fm$^{4}$.
Because the ratio $B(E2, 4^{+}_{6} \rightarrow 2^{+}_{6})/B(E2, 2^{+}_{6} \rightarrow 0^{+}_{5}) = 1.31$ 
is consistent with that for the rigid rotor model $10/7$, 
this band is considered to be the rotational band of the linear-chain structure with the $^{10}$Be+$\alpha$ correlation.
These linear-chain states may be the candidates for the states observed recently in $^{10}$Be+$\alpha$ breakup reactions.
When the energies relative to the $^{10}$Be+$\alpha$ threshold are considered,
these experimental states are observed in the energy region similar to the theoretical linear-chain band.
The significant branching ratios of $^{10}$Be+$\alpha$ decays of these states may support the $^{10}$Be+$\alpha$ cluster structure.
We should comment here that such a rigid rotational band of the $3\alpha$ linear-chain structure does not appear in $^{12}$C.
In the stabilization of the linear-chain structure in $^{14}$C, 
excess neutrons are expected to play an important role.
Detailed discussion of its mechanism is given in the next section.

Other states have no specific structure and are difficult to be classified as band members.
The observed $1^{+}$ states are missing in the present calculations, 
because spin-aligned configurations are unfavored; therefore 
such the states cannot be described in the $\beta$-$\gamma$ constraint variation 
where only the lowest energy solution is obtained at each deformation. 
To describe the low-lying $1^{+}$ state, it is necessary to introduce other constraints, 
for example, on the expectation value of intrinsic spin.

\begin{figure}[tb]
\centering
	\includegraphics[angle=-90,width=8.6cm, bb=290 54 542 392, clip]{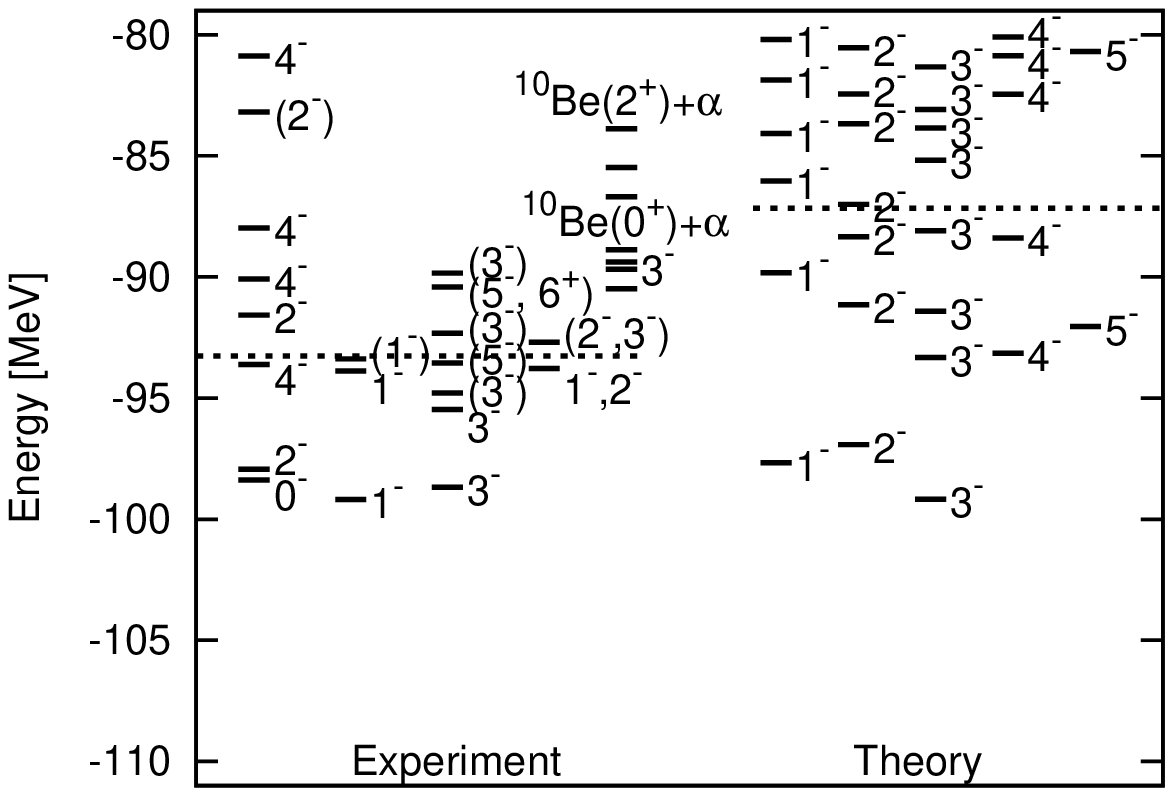}
	\caption{Energy levels of the negative-parity states in $^{14}$C.
		Five columns on the left are the experimental data and five columns on the right are the calculated results.
		The dotted lines on the left and right show 
		the experimental and theoretical $^{10}$Be$+ \alpha$ threshold energies, respectively.}
\label{Energy_level_14C_-}
\end{figure}

\begin{table}[tb]
	\caption{Electromagnetic transition strengths $B(E2)$ which are stronger than 1.0 $e^{2}$fm$^{4}$ for the negative-parity states in $^{14}$C.
	The unit is $e^{2}$fm$^{4}$.}
	\label{E2_14C_-}
	\begin{tabular}[t]{ccc} \hline \hline
\multirow{2}{*}{Transition} & \multicolumn{2}{c}{Strength} \\ \cline{2-3}
 & Theory & Experiment \\ \hline
$1^{-}_{2} \rightarrow 1^{-}_{1}$ &  5.9 & \\
$1^{-}_{3} \rightarrow 1^{-}_{2}$ &  2.1 & \\
$1^{-}_{4} \rightarrow 1^{-}_{3}$ &  1.4 & \\
$1^{-}_{5} \rightarrow 1^{-}_{2}$ &  3.5 & \\
$1^{-}_{5} \rightarrow 1^{-}_{3}$ &  1.3 & \\
$1^{-}_{5} \rightarrow 1^{-}_{4}$ & 15.5 & \\
$1^{-}_{6} \rightarrow 1^{-}_{3}$ &  3.0 & \\
$1^{-}_{6} \rightarrow 1^{-}_{4}$ & 11.5 & \\
$1^{-}_{6} \rightarrow 1^{-}_{5}$ & 19.2 & \\
 & & \\
$2^{-}_{2} \rightarrow 1^{-}_{2}$ &  1.3 & \\
$2^{-}_{3} \rightarrow 1^{-}_{1}$ &  2.2 & \\
$2^{-}_{3} \rightarrow 1^{-}_{2}$ &  1.6 & \\
$2^{-}_{3} \rightarrow 1^{-}_{4}$ &  3.5 & \\
$2^{-}_{4} \rightarrow 1^{-}_{3}$ &  1.0 & \\
$2^{-}_{5} \rightarrow 1^{-}_{3}$ &  1.1 & \\
$2^{-}_{5} \rightarrow 1^{-}_{4}$ &  8.3 & \\
$2^{-}_{5} \rightarrow 1^{-}_{5}$ &  3.2 & \\
$2^{-}_{6} \rightarrow 1^{-}_{4}$ & 20.5 & \\
$2^{-}_{6} \rightarrow 1^{-}_{5}$ &  2.7 & \\
$2^{-}_{7} \rightarrow 1^{-}_{3}$ &  1.9 & \\
$2^{-}_{7} \rightarrow 1^{-}_{5}$ &  1.7 & \\
$2^{-}_{7} \rightarrow 1^{-}_{6}$ &  1.3 & \\
 & & \\
$2^{-}_{3} \rightarrow 2^{-}_{2}$ &  1.4 & \\
$2^{-}_{4} \rightarrow 2^{-}_{1}$ &  1.1 & \\
$2^{-}_{5} \rightarrow 2^{-}_{2}$ &  1.1 & \\
$2^{-}_{6} \rightarrow 2^{-}_{5}$ &  1.9 & \\
$2^{-}_{7} \rightarrow 2^{-}_{3}$ &  1.6 & \\
$2^{-}_{7} \rightarrow 2^{-}_{6}$ &  4.6 & \\
	\hline \hline
	\end{tabular} \qquad
	\begin{tabular}[t]{ccc} \hline \hline
\multirow{2}{*}{Transition} & \multicolumn{2}{c}{Strength} \\ \cline{2-3}
 & Theory & Experiment \\ \hline
$3^{-}_{1} \rightarrow 1^{-}_{1}$ &  1.3 & $3.0 \pm 1.2$ \\
$3^{-}_{2} \rightarrow 1^{-}_{1}$ &  1.9 & \\
$3^{-}_{4} \rightarrow 1^{-}_{1}$ &  1.2 & \\
$3^{-}_{4} \rightarrow 1^{-}_{4}$ &  2.1 & \\
$3^{-}_{5} \rightarrow 1^{-}_{4}$ &  1.2 & \\
$3^{-}_{6} \rightarrow 1^{-}_{4}$ & 10.1 & \\
$3^{-}_{7} \rightarrow 1^{-}_{4}$ &  2.4 & \\
$3^{-}_{7} \rightarrow 1^{-}_{6}$ &  1.3 & \\
$3^{-}_{8} \rightarrow 1^{-}_{2}$ &  3.2 & \\
$3^{-}_{8} \rightarrow 1^{-}_{3}$ &  3.3 & \\
$3^{-}_{8} \rightarrow 1^{-}_{4}$ &  1.1 & \\
$3^{-}_{8} \rightarrow 1^{-}_{5}$ & 10.0 & \\
$3^{-}_{8} \rightarrow 1^{-}_{6}$ &  5.1 & \\
 & & \\
$3^{-}_{1} \rightarrow 2^{-}_{1}$ &  1.4 & \\
$3^{-}_{2} \rightarrow 2^{-}_{2}$ &  1.5 & \\
$3^{-}_{2} \rightarrow 2^{-}_{3}$ &  1.4 & \\
$3^{-}_{3} \rightarrow 2^{-}_{1}$ &  4.1 & \\
$3^{-}_{6} \rightarrow 2^{-}_{3}$ &  2.3 & \\
$3^{-}_{6} \rightarrow 2^{-}_{6}$ &  5.6 & \\
$3^{-}_{7} \rightarrow 2^{-}_{2}$ &  1.7 & \\
$3^{-}_{7} \rightarrow 2^{-}_{5}$ &  1.4 & \\
$3^{-}_{7} \rightarrow 2^{-}_{6}$ &  1.5 & \\
$3^{-}_{8} \rightarrow 2^{-}_{2}$ &  1.0 & \\
$3^{-}_{8} \rightarrow 2^{-}_{3}$ &  3.9 & \\
$3^{-}_{8} \rightarrow 2^{-}_{6}$ &  5.1 & \\
$3^{-}_{8} \rightarrow 2^{-}_{7}$ & 11.2 & \\
 & & \\
$3^{-}_{2} \rightarrow 3^{-}_{1}$ &  3.3 & \\
$3^{-}_{3} \rightarrow 3^{-}_{2}$ &  1.1 & \\
$3^{-}_{4} \rightarrow 3^{-}_{1}$ &  1.1 & \\
$3^{-}_{5} \rightarrow 3^{-}_{2}$ &  1.2 & \\
$3^{-}_{5} \rightarrow 3^{-}_{4}$ &  1.2 & \\
$3^{-}_{6} \rightarrow 3^{-}_{4}$ &  1.4 & \\
$3^{-}_{6} \rightarrow 3^{-}_{5}$ &  2.5 & \\
$3^{-}_{7} \rightarrow 3^{-}_{6}$ &  5.9 & \\
	\hline \hline
	\end{tabular} \qquad
	\begin{tabular}[t]{ccc} \hline \hline
\multirow{2}{*}{Transition} & \multicolumn{2}{c}{Strength} \\ \cline{2-3}
 & Theory & Experiment \\ \hline
$4^{-}_{2} \rightarrow 2^{-}_{1}$ &  1.6 & \\
$4^{-}_{3} \rightarrow 2^{-}_{3}$ &  4.7 & \\
$4^{-}_{3} \rightarrow 2^{-}_{5}$ &  4.2 & \\
$4^{-}_{3} \rightarrow 2^{-}_{6}$ &  6.9 & \\
$4^{-}_{4} \rightarrow 2^{-}_{2}$ &  1.2 & \\
$4^{-}_{5} \rightarrow 2^{-}_{6}$ & 13.7 & \\
 & & \\
$4^{-}_{1} \rightarrow 3^{-}_{1}$ &  4.8 & \\
$4^{-}_{1} \rightarrow 3^{-}_{2}$ &  2.0 & \\
$4^{-}_{2} \rightarrow 3^{-}_{3}$ &  3.2 & \\
$4^{-}_{3} \rightarrow 3^{-}_{4}$ &  8.4 & \\
$4^{-}_{3} \rightarrow 3^{-}_{6}$ &  3.9 & \\
$4^{-}_{3} \rightarrow 3^{-}_{7}$ &  1.1 & \\
$4^{-}_{3} \rightarrow 3^{-}_{8}$ &  1.5 & \\
$4^{-}_{4} \rightarrow 3^{-}_{5}$ &  1.3 & \\
$4^{-}_{4} \rightarrow 3^{-}_{7}$ &  2.8 & \\
$4^{-}_{5} \rightarrow 3^{-}_{5}$ &  1.2 & \\
$4^{-}_{5} \rightarrow 3^{-}_{6}$ & 12.4 & \\
$4^{-}_{5} \rightarrow 3^{-}_{7}$ &  2.6 & \\
$4^{-}_{5} \rightarrow 3^{-}_{8}$ &  4.0 & \\
 & & \\
$4^{-}_{5} \rightarrow 4^{-}_{3}$ &  1.6 & \\
$4^{-}_{5} \rightarrow 4^{-}_{4}$ &  1.5 & \\
 & & \\
$5^{-}_{1} \rightarrow 3^{-}_{1}$ &  3.9 & \\
$5^{-}_{1} \rightarrow 3^{-}_{2}$ &  1.1 & \\
$5^{-}_{2} \rightarrow 3^{-}_{4}$ &  2.2 & \\
$5^{-}_{2} \rightarrow 3^{-}_{6}$ & 14.3 & \\
$5^{-}_{2} \rightarrow 3^{-}_{7}$ &  5.0 & \\
 & & \\
$5^{-}_{1} \rightarrow 4^{-}_{1}$ &  2.8 & \\
$5^{-}_{2} \rightarrow 4^{-}_{3}$ &  1.8 & \\
$5^{-}_{2} \rightarrow 4^{-}_{5}$ &  1.3 & \\
	\hline \hline
	\end{tabular}
\end{table}

Next, we describe the results for the negative-parity states.
We show the calculated energy levels of the negative-parity states in Fig.~\ref{Energy_level_14C_-}
as well as the experimental levels.
In the four columns from the left, we display the experimental energy levels of the negative-parity assigned states \cite{Ajzenberg_A=13-15_91}.
In the fifth column, we show the energy levels recently observed 
in $^{10}$Be+$\alpha$ breakup reactions \cite{Milin_14C_04, Soic_CandBisotopes_04, Price_14C_07, Haigh_14C_08}.
The theoretical levels are illustrated in each spin.
The calculated $E2$ transition strengths are listed in Table~\ref{E2_14C_-} along with experimental data \cite{Ajzenberg_A=13-15_91}.

The calculated low-lying states have large overlaps with the basis wave functions in the small deformation region.
For instance, the $3^{-}_{1}$ state has 87\% overlap with $(\beta \cos \gamma, \beta \sin \gamma) = (0.20, 0.09)$ (Fig.~\ref{density_14C_-}(b)),
which is the minimum point in the $3^{-}$ energy surface.
As mentioned before, this state is interpreted as the intermediate between the cluster structure and 
the shell-model structure with the almost $p_{3/2}$-subshell closed-proton configuration and
the $1p$-$1h$ excited neutron configuration.
The calculated strength, $B(E2, 3^{-}_{1} \rightarrow 1^{-}_{1})=1.3$ $e^{2}$fm$^{4}$ is reasonable compared with 
the experimental value, $B(E2, 3^{-}_{1} \rightarrow 1^{-}_{1})=3.0 \pm 1.2$ $e^{2}$fm$^{4}$,
though the level ordering is somehow in disagreement with the experimental one.
As seen in the $E2$ transitions in Table~\ref{E2_14C_-}, 
$E2$ transitions among the low-lying states below the $^{10}$Be$+ \alpha$ threshold are not strong and 
they show no remarkable collectivity nor specific band assignments for these states.

\begin{figure}[tb]
\centering
	\includegraphics[angle=-90,width=8.6cm, bb=160 53 542 392, clip]{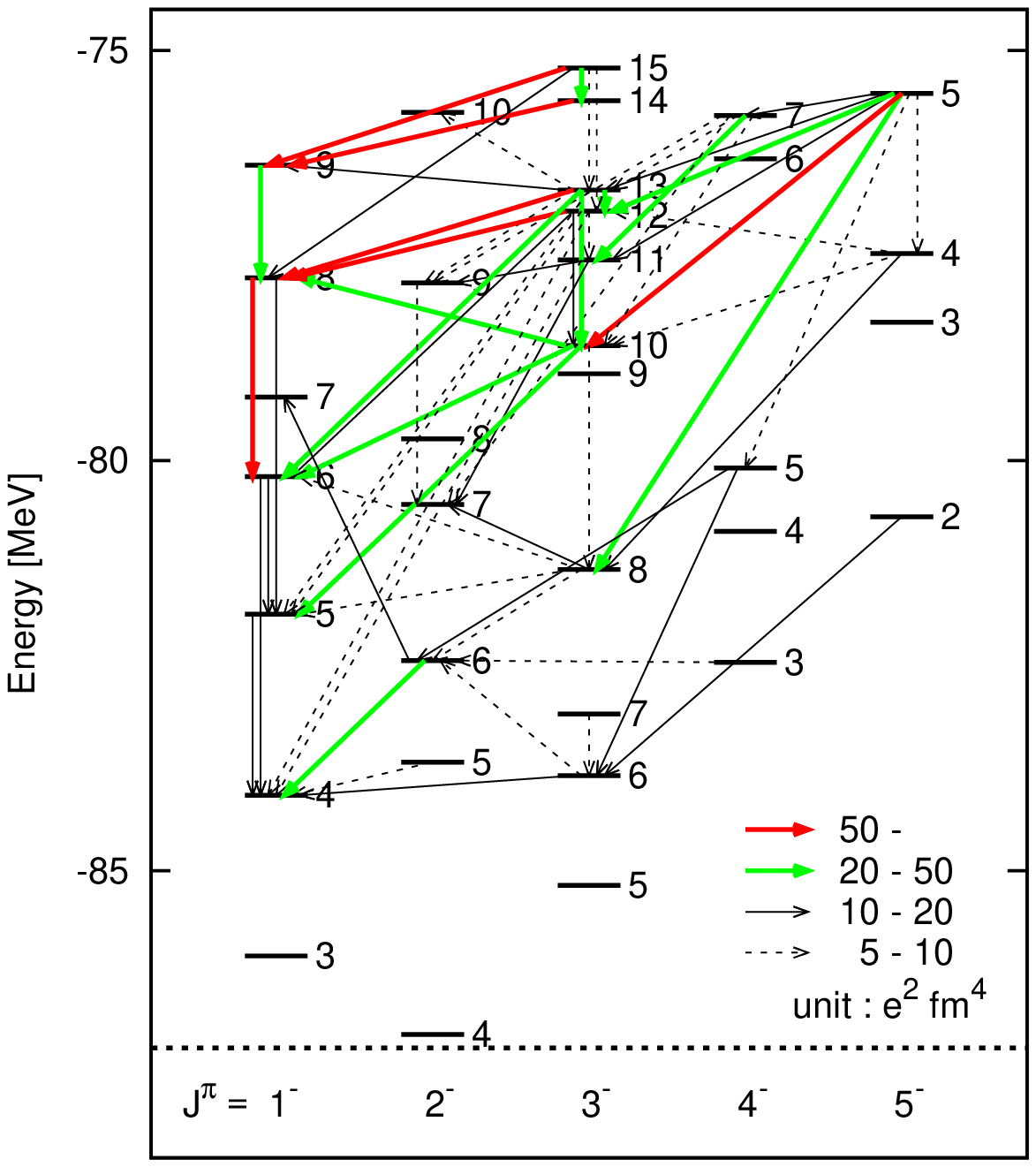}
	\caption{(Color online) Calculated energy levels of the negative-parity states above the $^{10}$Be$+ \alpha$ threshold energy 
		with $E2$ transition strengths in $^{14}$C.
		An integer $i$ at the right of a level denotes the order of $J^{\pi}_{i}$ states.
		$5.0$ $e^{2}$fm$^{4} < B(E2) \le 10.0$ $e^{2}$fm$^{4}$,
		$10.0$ $e^{2}$fm$^{4} < B(E2) \le 20.0$ $e^{2}$fm$^{4}$, 
		$20.0$ $e^{2}$fm$^{4} < B(E2) \le 50.0$ $e^{2}$fm$^{4}$, and
		$50.0$ $e^{2}$fm$^{4} < B(E2)$ transitions are described by 
		broken, black solid, green bold solid, and red bold solid arrows, respectively.
		The dotted line shows theoretical $^{10}$Be$+ \alpha$ threshold energy.}
\label{Energy_level_14C_-_E2}
\end{figure}

In the highly excited negative-parity states above the $^{10}$Be$+ \alpha$ threshold energy, 
we obtain many developed cluster states having significant overlaps with the basis wave functions in the large $\beta$ regions.
Some strong $E2$ transitions are found in the calculated $B(E2)$ values for high-lying states.
However, the $E2$ strengths are fragmented into many states; therefore, 
it is difficult to classify these states in simple band assignments.
In Fig.~\ref{Energy_level_14C_-_E2}, 
we show the calculated level scheme of the high-lying negative-parity states 
above the $^{10}$Be$+ \alpha$ threshold energy with $E2$ transition strengths.
Among these highly excited states, some flows of strong $E2$ transition strengths are found in the present calculations.
Transitions from each state are, however, dispersed.
For example, the $5^{-}_5$ state has strong transitions to the $3^{-}_{8}$, $3^{-}_{10}$, and $3^{-}_{12}$ states.
These states are constituted by largely deformed states such as those shown in Figs.~\ref{density_14C_-}(c), (d), and (e).
What is different in the linear-chain band in the positive-parity states
is that the GCM amplitudes of these negative-parity states do not concentrate on a single basis wave function 
with a specific geometric configuration but distribute into many basis wave functions.
These features indicate strong mixing of the largely deformed prolate, oblate, and triaxial states.
This mixing may be an origin of the complicated level scheme of the negative-parity states. 
As a result, the parity partner of the linear-chain band suggested in the positive-parity states 
may disappear in the negative-parity states.
In spite of the largely scattered $E2$ strengths,
there are sets of states classified by flows of significant strong $E2$ transitions.
For example, a set consisting of $5^{-}_5$, $3^{-}_{10}$, $3^{-}_{12}$, $3^{-}_{13}$, $1^{-}_{5}$, $1^{-}_{6}$, and $1^{-}_{8}$ states 
can be interpreted as members of a quasiband with $^{10}$Be+$\alpha$ cluster structure in the negative-parity states of $^{14}$C.
In the next section, we explain a possible reason why this mixing occurs.

We here comment the possible reason for missing of the low-lying $0^{-}$ state in the present calculation. 
It is expected to have $1p$-$1h$ configuration with one neutron in the $1s_{1/2}$ orbital. 
In the present framework, the width parameter is taken to be a common value 
for all the single-particle Gaussian wave functions. 
The fixed Gaussian wave packets may not be suitable to describe the $1s_{1/2}$ orbital 
because the spatial extent of the $1s_{1/2}$ orbital should be large compared with other orbitals.

\section{Discussion}\label{discussion}

In this section, we discuss the linear-chain band and the role of the excess neutrons.
We also discuss the cluster structures of $^{14}$C in comparison with the findings of earlier works.

\begin{figure}[tb]
	\begin{tabular}{cc}
	\includegraphics[width=6.1cm, bb=7 11 592 301, clip]{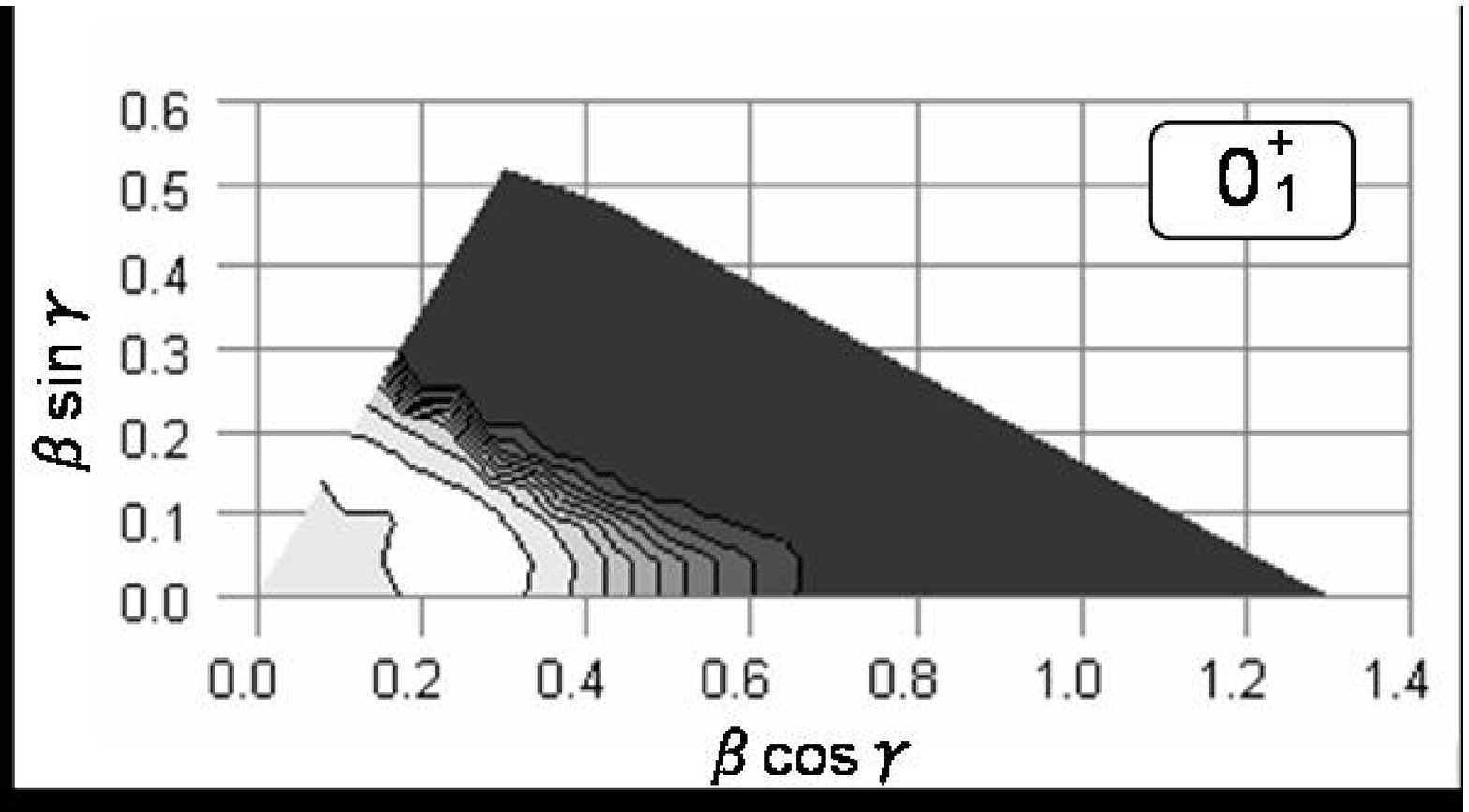} & \includegraphics[width=0.85cm, bb=27 2 66 165, clip]{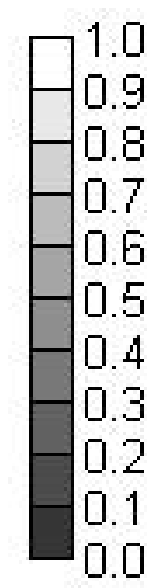} \\
	\includegraphics[width=6.1cm, bb=7 11 592 301, clip]{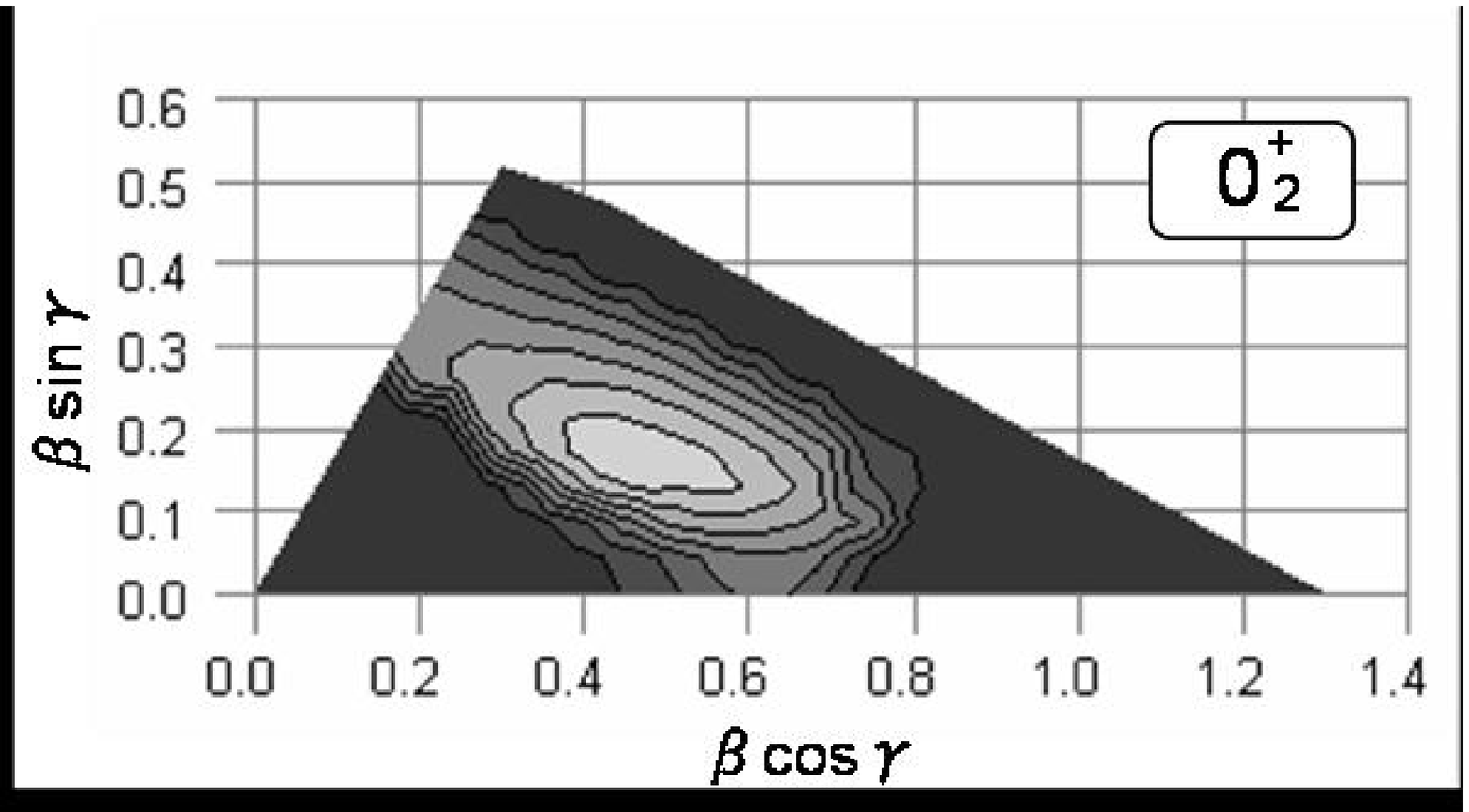} & \\
	\includegraphics[width=6.1cm, bb=7 11 592 301, clip]{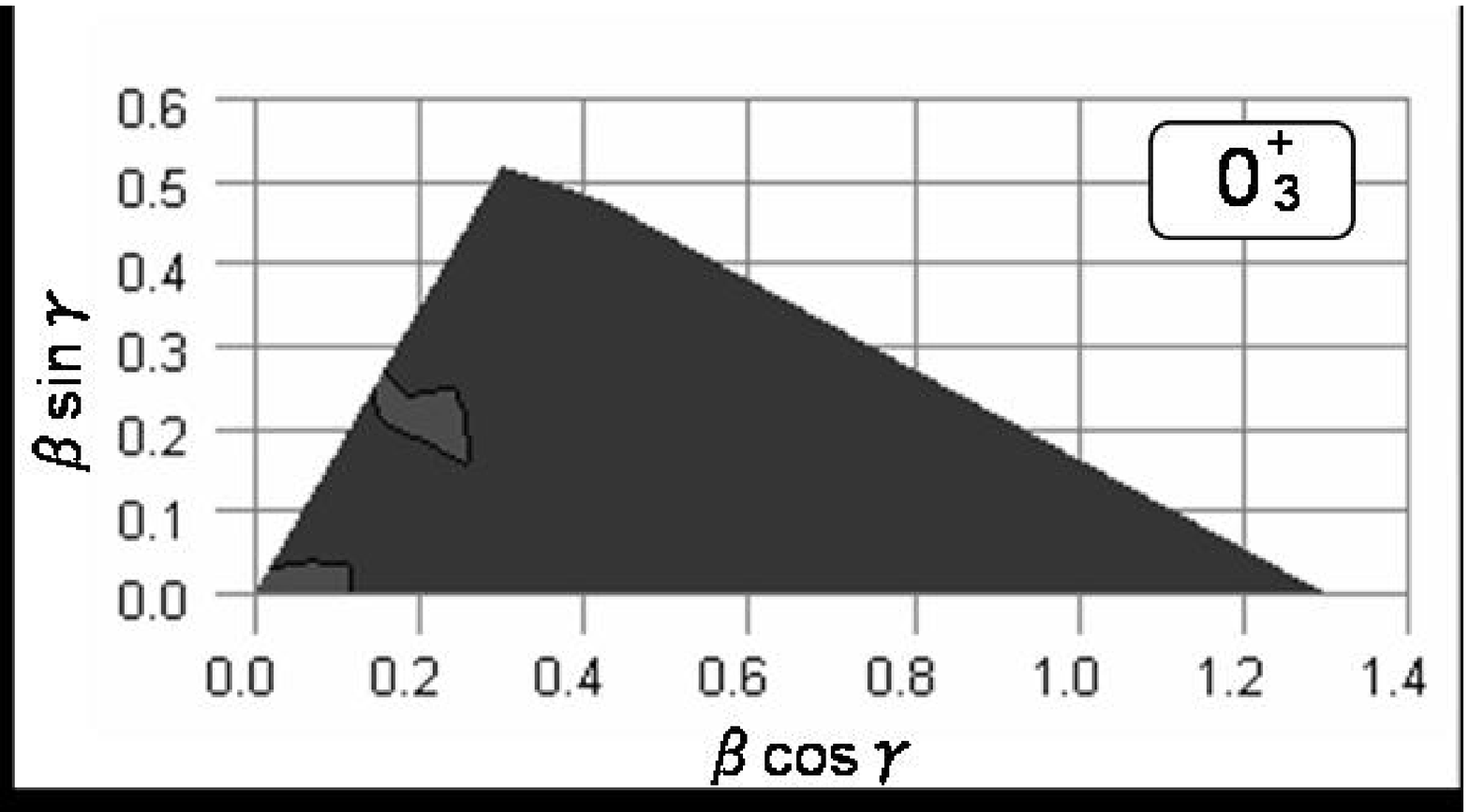} & \\
	\includegraphics[width=6.1cm, bb=7 11 592 301, clip]{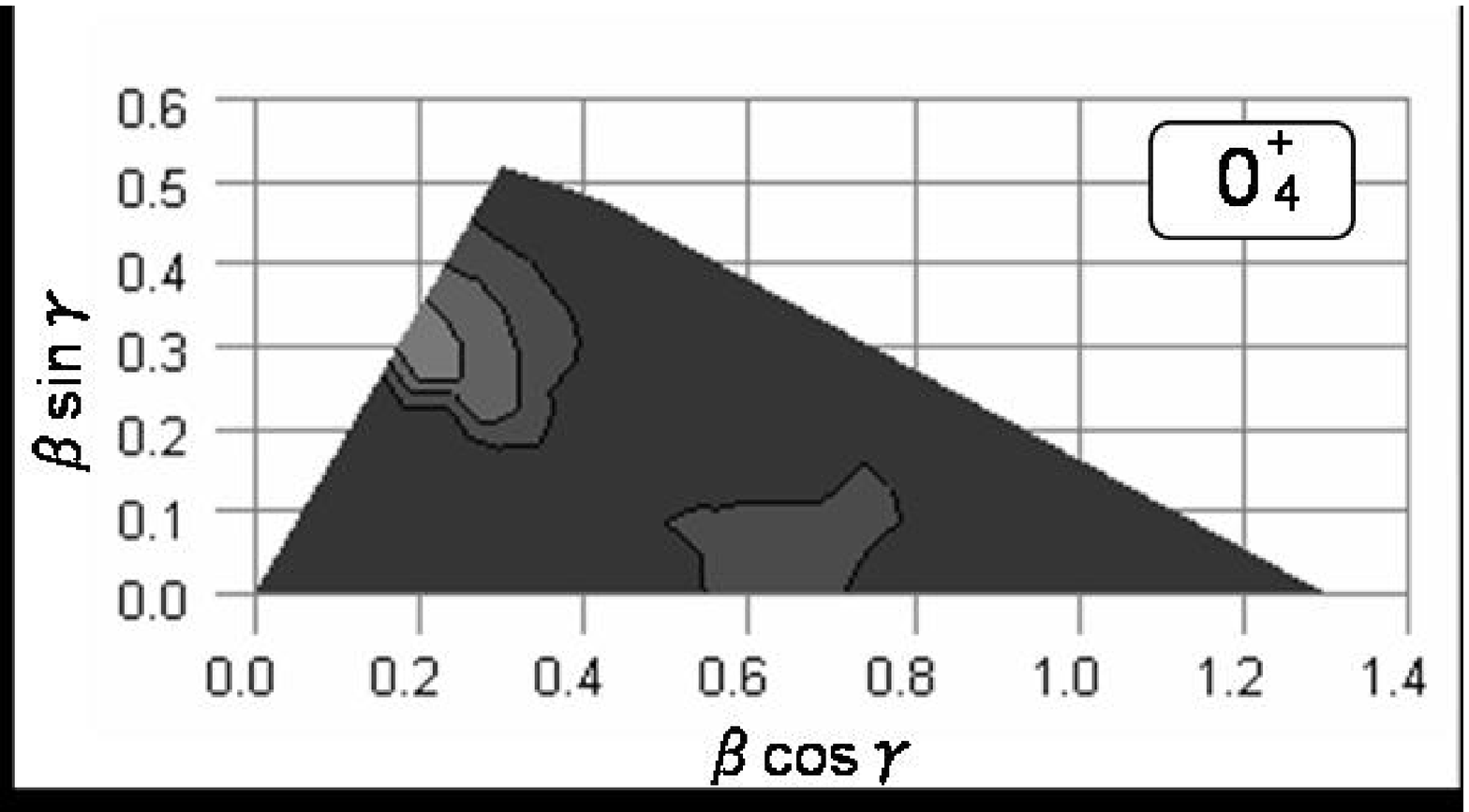} & \\
	\includegraphics[width=6.1cm, bb=7 11 592 301, clip]{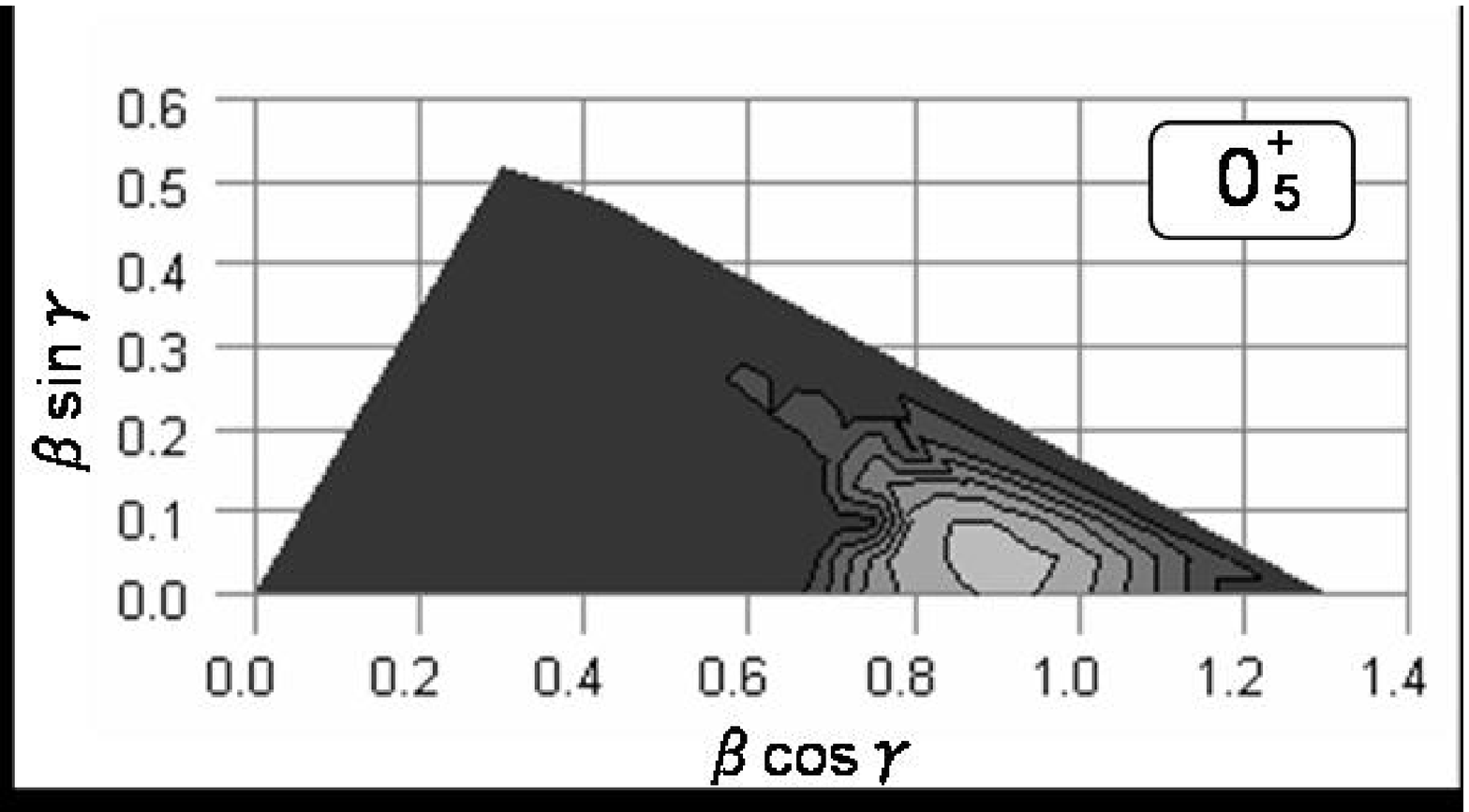} & \\
	\includegraphics[width=6.1cm, bb=7 11 592 301, clip]{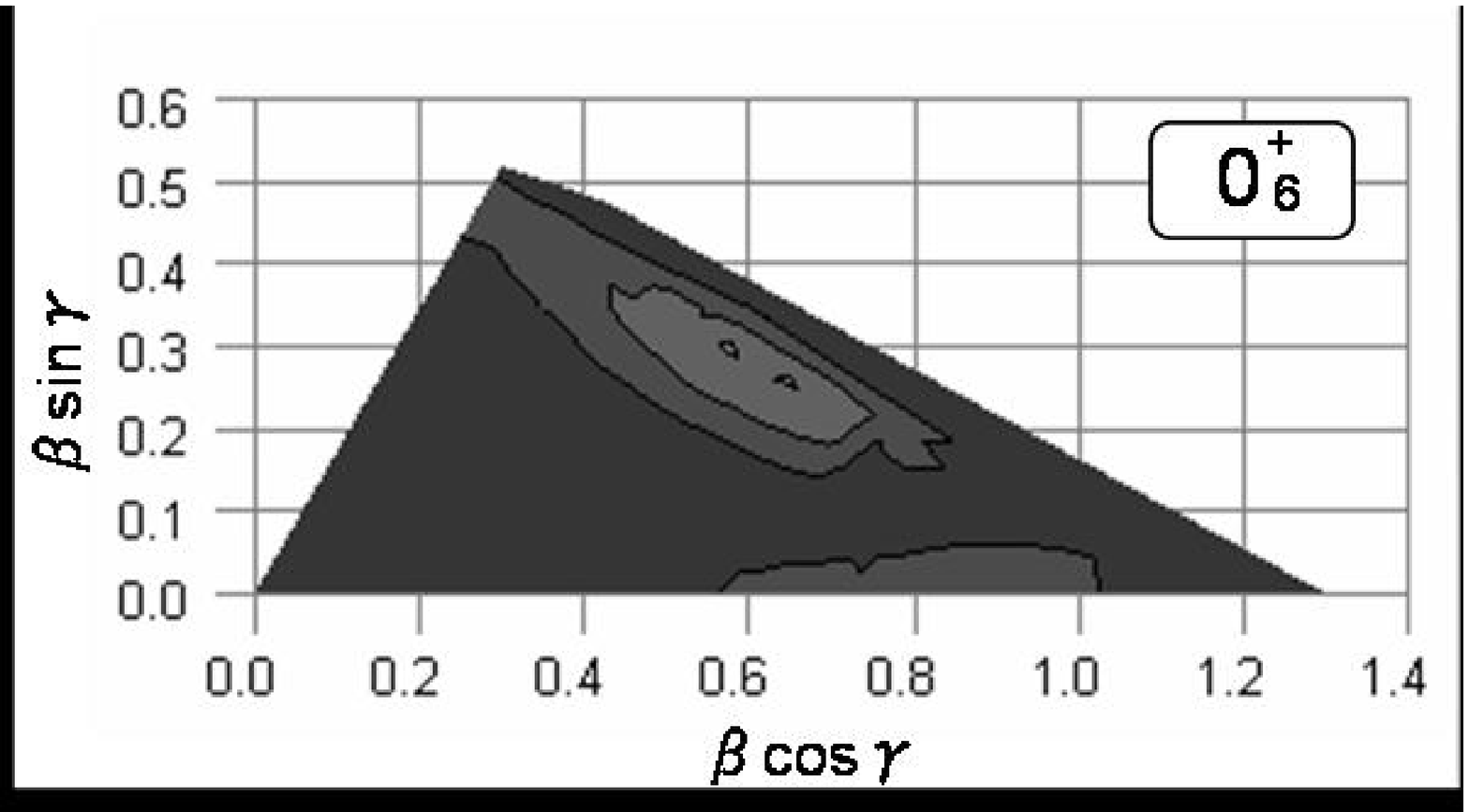} & \\
	\end{tabular}
	\caption{Overlaps of the GCM wave functions for the $0^{+}$ states of $^{14}$C with 
		basis AMD wave functions, $\hat{P}^{J=0}_{00} | \Phi^+ (\beta,\gamma) \rangle$, obtained with the $\beta$-$\gamma$ constraint.
		The overlap at a certain ($\beta,\gamma$) point is shown in the contour map.}
	\label{14C_overlap_0+}
\end{figure}

First, we discuss the reasons why the linear-chain band appears in the positive-parity states of $^{14}$C. 
For appearance of the linear-chain band, 
both the existence of the plateau around the linear-chain structure in the energy surface 
and the orthogonality to the lower states are essential.
To explain more details, we show the squared overlaps 
$|\langle 0^+| \hat{P}^{J=0}_{00} | \Phi^+ (\beta,\gamma) \rangle|^2$
of the GCM wave functions for the $0^{+}$ states with 
the basis AMD wave function at each point $(\beta, \gamma)$ in Fig.~\ref{14C_overlap_0+}.
As mentioned before, 
the flat region exists near the largely deformed prolate region in the energy surfaces, and there exists a local minimum.
Although this is a shallow local minimum with 1 MeV depth in the $0^{+}$ energy surface, 
the GCM amplitudes for the linear-chain $0^{+}_{5}$ state localize around this local minimum as shown in Fig.~\ref{14C_overlap_0+}.
Another reason is the orthogonality to the triaxial bands consisting of the  $K^{\pi}=0^{+}_{2}$ band and its $K^{\pi}=2^{+}$ side band.
In the case of $0^{+}$ states, the GCM amplitudes for the $0^{+}_{2}$ state occupy the triaxial region. 
We remind the reader that the bending $3\alpha$ configurations are contained in the triaxial region.
Therefore, even though the energy curve around the linear-chain structure is soft against the bending mode,  
the GCM amplitudes for the linear-chain $0^{+}_{5}$ state are confined in the prolate region 
to satisfy the orthogonality to the triaxial $0^{+}_{2}$ state resulting 
in the concentration of the amplitude at the linear-chain structure. 
In other words, the triaxial $0^{+}_{2}$ state prevents the linear-chain $0^{+}_{5}$ state from bending.
The situation is almost the same in the $2^{+}$ and $4^{+}$ states except that there are non-zero $K$ components.
In the $2^{+}$ case, distributions of the squared overlap on the $\beta$-$\gamma$ plane for the $2^{+}_{2}$ and $2^{+}_{6}$ states are
similar to the $0^{+}_{2}$ and $0^{+}_{5}$ states, respectively. 
Moreover, the distribution of the squared overlap for the $2^{+}_{3}$ state is similar to those for the $2^{+}_{2}$ state.
This means the $2^{+}_{2}$ and $2^{+}_{3}$ states already exhaust the $K^{\pi}=0^{+}$ and $2^{+}$ components
of the bases which have bending configuration.
Therefore, the $2^{+}$ states can be discussed in the same line as the $0^{+}$ states.
That is, in the stabilization of the linear-chain state, 
the orthogonality of the $2^{+}_{6}$ state to the $2^{+}_{2}$ and $2^{+}_{3}$ states plays the same role as 
that of the $0^{+}_{5}$ state to the $0^{+}_{2}$ state.
The $2^{+}_{2}$ and $2^{+}_{3}$ states prevent the linear-chain $2^{+}_{6}$ state from bending,
so that the GCM amplitudes for the $2^{+}_{6}$ state concentrate around the linear-chain structure.

\begin{figure}[tb]
	\includegraphics[width=8.6cm, bb=10 12 700 364, clip]{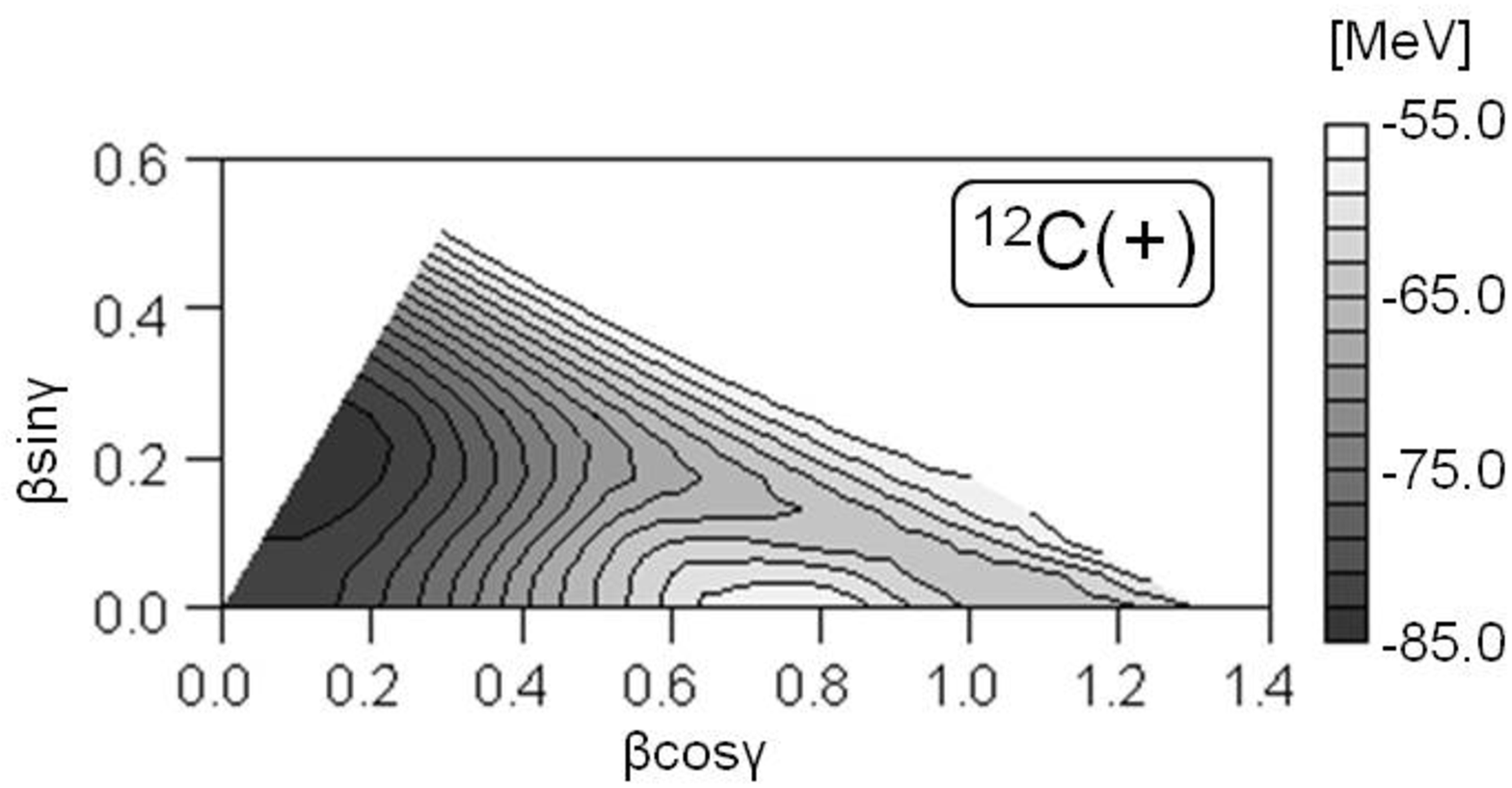} \\
	\includegraphics[width=8.6cm, bb=10 12 700 364, clip]{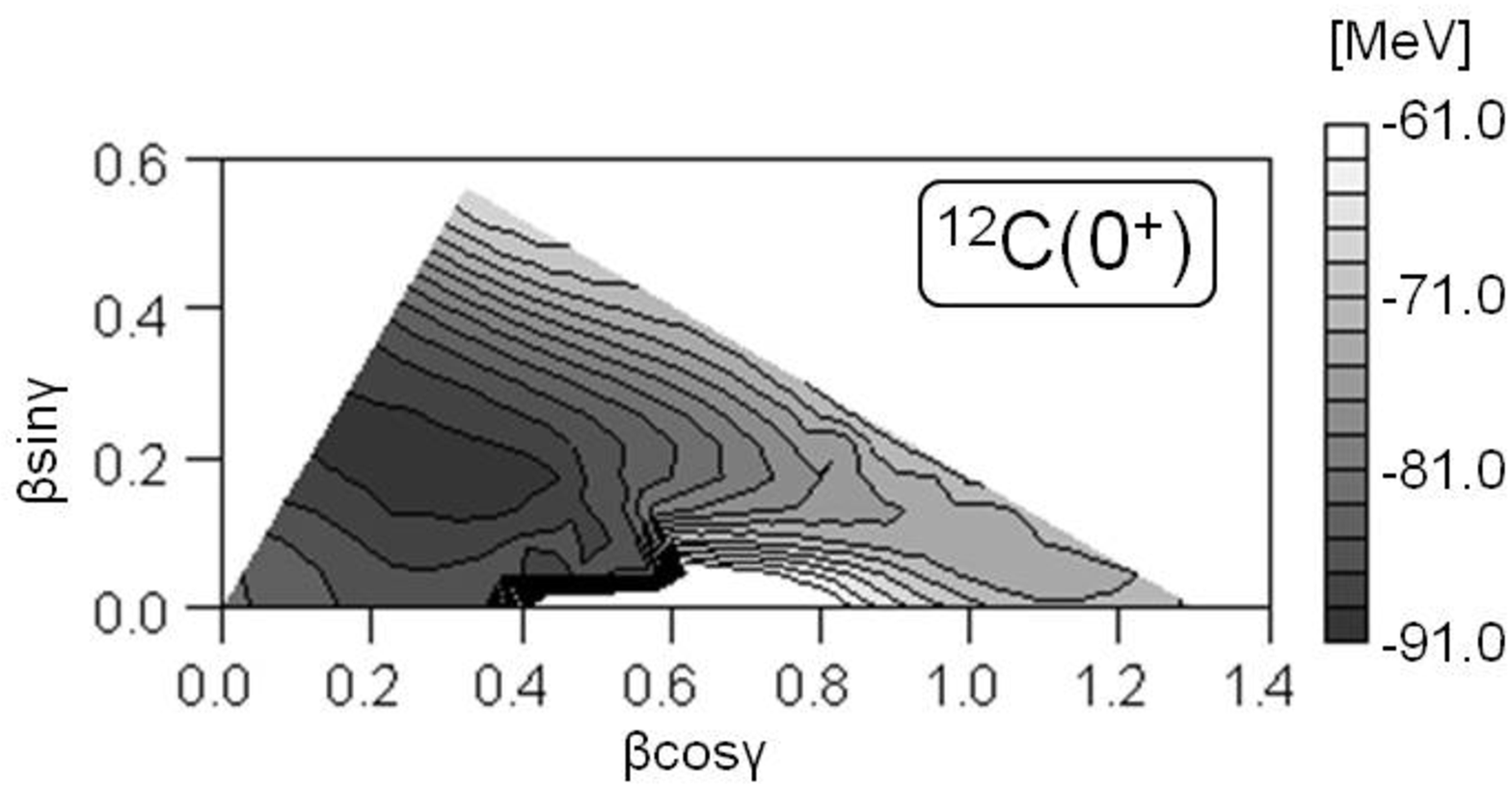}
	\caption{Energy surfaces of $^{12}$C on the $\beta$-$\gamma$ plane.
	The upper shows the energy for the positive-parity states before the total-angular-momentum projection and
	the lower shows that for the $0^{+}$ states after the total-angular-momentum projection.}
	\label{12C_energy_surface_+}
\end{figure}

Next, we consider the role of the excess neutrons in the linear-chain structure of $^{14}$C. 
The linear-chain $3\alpha$ structure in C isotopes has been a long standing problem, 
for example, as discussed in Refs.~\cite{Morinaga_4N_56,Morinaga_12C_linear_chain_66,vonOertzen_Cisotopes_97,Itagaki_Cisotopes_01}. 
In the recent works on $^{12}$C with AMD and FMD calculations \cite{Neff_12C_04,En'yo_12C_07, Suhara_AMD_10}, 
an open triangle $3\alpha$ structure was suggested only in the $0^{+}_{3}$ state, 
but no straight-line 3$\alpha$-chain structure nor linear-chain band was found in the $^{12}$C system.
It is contrastive to $^{14}$C where the straight linear-chain structure is suggested 
in the $0^{+}_{5}$, $2^{+}_{6}$, $4^{+}_{6}$ states in the present work.
It means that there is no linear-chain state in $^{12}$C except for the $0^{+}_{3}$ state,
while the linear-chain structure construct the rotational band in $^{14}$C.
By comparing the present results of $^{14}$C with the $^{12}$C results 
calculated with the same framework of the $\beta$-$\gamma$ constraint AMD + GCM in Ref.~\cite{Suhara_AMD_10}, 
it is found that the difference between the $^{14}$C and the $^{12}$C cases 
originates in the differences of the behavior of the energy surfaces and 
also the structure of the states lower than the linear-chain structure.
First, we describe the difference in the energy surfaces.
In Fig.~\ref{12C_energy_surface_+}, we show the energy surfaces for $^{12}$C 
to compare them with the those for $^{14}$C.
The detailed discussions for $^{12}$C are written in Ref.~\cite{Suhara_AMD_10}.
In the energy surfaces of $^{12}$C (Fig.~9), 
there is a flat region around the largely deformed prolate region
but there is no local minimum even in the $0^{+}$ energy surface.
However, in the energy surfaces of $^{14}$C (Fig.~1), 
there is a shallow local minimum with 1 MeV depth around the largely deformed prolate region.
This difference indicates that excess neutrons give the effect to favor the linear-chain structure.
Second, we discuss the structure of the lower states than the linear-chain structure.
As explained in the previous paragraph, existence of the triaxial bands 
is essential for the linear-chain band in $^{14}$C because they prevent 
the linear-chain band from bending because of the orthogonality.
However, in $^{12}$C, the states which prevent the linear-chain states from bending do not exist. 
An exception is the case of $0^{+}$ states.
Let us first explain the $0^{+}$ cases, which are similar in $^{12}$C and $^{14}$C.
In $^{12}$C, GCM amplitudes of the $0^{+}_{2}$ state occupy the considerably bending $3\alpha$ structure 
and distribute widely in various basis wave functions.
Therefore, the $0^{+}_{2}$ state prevents the $0^{+}_{3}$ state from bending extremely
because of the orthogonality.
Because the lower state, $0^{+}_{2}$, already exhausts the bases which have bending configuration, 
the higher state, $0^{+}_{3}$, cannot have bending configuration; therefore, the linear-chainlike structure dominates 
the $0^{+}_{3}$ state even though the linear-chain structure is not a local minimum in the energy surface in $^{12}$C.
This is similar to the relation between the $0^{+}_{2}$ and $0^{+}_{5}$ states in $^{14}$C.
However, for the $2^{+}$ and $4^{+}$ states in $^{12}$C,
there is no such a state that exhausts bases with bending configurations.
Therefore, mixing of the largely deformed prolate, oblate and triaxial states occurs.
To summarize, the excess neutrons play two roles in the appearance of the rotational band of the linear-chain structure. 
First, because of the existence of the excess neutrons, 
the energy surfaces for the positive-parity states change to favor the linear-chain structure. 
Second, the excess neutrons are essential for constructing the triaxial bands, 
which occupy the bending $3\alpha$ configurations considerably and prevent
the linear-chain states from bending to satisfy the orthogonality to each other.

As we explain in the previous section, the linear-chain band does not appear in the negative-parity states.
The reason is clear from the behavior of the energy surfaces for the negative-parity states.
There is no flat region in the largely deformed prolate region.
This means the $^{10}$Be cluster rotates easily in the negative-parity states.
This nature was already explained also in the discussion of the intrinsic density distributions shown in Fig.~\ref{density_14C_-}(e).
Because of this unstable feature against rotational mode of the $^{10}$Be cluster, 
the prolate deformed states strongly mix with the triaxially deformed states in the negative-parity states.

Linear-chain structures of C isotopes have been discussed 
in earlier works with 3$\alpha$+$Xn$ cluster models \cite{vonOertzen_Cisotopes_97,Itagaki_Cisotopes_01,vonOertzen_14C_04}.
In the study by Itagaki {\it et al.} \cite{Itagaki_Cisotopes_01}, 
it was reported that the linear-chain state may not be stable in $^{14}$C. 
Their argument seems to contradict the present results of the linear-chain band suggested in the positive-parity states of $^{14}$C.
In their work, it was shown that the linear-chain structure of $^{14}$C is unstable against the bending mode 
because there is no local minimum in the energy curve. 
However, to confirm the stability of an excited state, 
it is essential to take into account the orthogonality to lower states as done in the GCM calculations. 
In the results of Ref.~\cite{Itagaki_Cisotopes_01}, the energy curve along the bending mode is very flat 
and is not so much different from the present results though it has no local minimum in their calculations 
while the shallow local minimum exists in the flat region of the present results. 
As mentioned before, the linear-chain band appears from this flat region mainly 
because of the orthogonality to the triaxial bands which block the bending of the $3\alpha$ configuration. 
Therefore, we should stress again the importance of the GCM calculations for excited states.
Another difference of the model used in Ref.~\cite{Itagaki_Cisotopes_01} from the preset framework 
is that three $\alpha$ clusters are set at the same intervals 
and the excess neutrons are assumed to occupy the orbital moving around the whole 3$\alpha$ core. 
In such a model, the system has to have a parity symmetric structure 
and the $^{10}$Be correlation cannot be incorporated. 
In the present calculations, though the 3$\alpha$ core is not {\it a priori} assumed in the framework, 
various 3$\alpha$+$2n$ structures are obtained in the energy variation. 
The geometric configuration of three $\alpha$ clusters and Gaussian centers for excess neutrons are optimized.
As a result, the linear-chain states show the $^{10}$Be correlation as discussed before. 
In further detailed analysis of the linear-chain structure, 
we confirm that the $^{10}$Be correlation gains energy compared with 
the symmetric configuration of the linear-chain 3$\alpha$+$2n$ structure.
These will be reported in a future article.

The idea of $^{10}$Be+$\alpha$ structure in $^{14}$C has been proposed
by von Oertzen {\it et al.} in Ref.~\cite{vonOertzen_14C_04}.
They conjectured that the parity doublet band could be built 
from the parity-asymmetric linear-chain structure because of the $^{10}$Be correlation.
However, the present results suggest that the linear-chain structure gives 
only the positive-parity rotational band but no negative-parity band.
As described before, because the $^{10}$Be core can easily rotate in the negative-parity $^{10}$Be+$\alpha$ states, 
the strong mixing of the linear-chain structure with various bending $^{10}$Be+$\alpha$ configurations occurs;
therefore, the parity partner of the positive-parity linear-chain band disappears.
Another difference between the present results and their suggestion is the energy position of the positive-parity linear-chain band. 
The linear-chain band is obtained above the $^{10}$Be+$\alpha$ threshold in the present calculations, 
whereas they proposed the linear-chain band below the threshold.
The reason for the higher energy than the $^{10}$Be+$\alpha$ threshold 
can be naturally understood by kinetic energy loss to align $^{10}$Be and $\alpha$ clusters in the linear-chain configuration. 

Recently, Itagaki {\it et al.} have studied excited states of $^{14}$C with a $3\alpha$+$2n$ cluster model and
have predicted the equilateral-triangular structure of the developed 
three $\alpha$ clusters with excess neutrons \cite{Itagaki_14C_04}.
They have proposed that the excess neutrons moving in molecular orbitals around the three $\alpha$ clusters 
stabilize the geometric $3\alpha$ configuration of equilateral triangle 
and suggested two rotational bands, $K^{\pi} = 0^{+}_{2}$ and $K^{\pi}=3^{-}$, resulting 
from the equilateral triangle intrinsic structure.
For the positive-parity states, we also obtained the $K^{\pi}=0^{+}_{2}$ band 
with well-developed three $\alpha$-cluster cores in an almost equilateral-triangular configuration, 
and therefore, in a sense, the present results seem to be partially consistent with the work of Itagaki {\it et al.}. 
However, the present results suggest the triaxiality owing to the excess neutron orbitals 
and the resultant $K^{\pi}=2^{+}$ side band on the $K^{\pi}=0^{+}_{2}$ band.
These are considerably different from the earlier work.
In addition, the development of three $\alpha$ clusters is weaker than that suggested in Ref.~\cite{Itagaki_14C_04} 
as the root-mean-square radius for the $0^{+}_{2}$ state is 2.45 fm in the present calculation 
while it is about 2.6 fm in the earlier work.
For the negative-parity states, a $K^{\pi}=3^{-}$ rotational band starting from 
the $3^{-}_{2}$ state was suggested in Ref.~\cite{Itagaki_14C_04}.
However, any clear $K^{\pi}=3^{-}$ rotational band cannot be found in the present calculations.
As mentioned, our results suggest that the $\alpha$ breaking 
is rather important in the low-lying states of $^{14}$C. 
The description of the $\alpha$ breaking is one of the advantages of the present model, 
while molecular orbitals for excess neutrons incorporated by Itagaki {\it et al.} 
are not completely included in our model. 
It is requested to investigate $^{14}$C in more details with a model including 
both of the $\alpha$ breaking and molecular orbitals around a clustering core.

\section{Summary and outlook}\label{summary}

We investigated structures of the ground and excited states of $^{14}$C with the method of $\beta$-$\gamma$ constraint AMD + GCM.
The results reproduce well observed data, such as the energy levels except for $1^{+}$ and $0^{-}$ states, 
the root-mean-square radii, and the $E2$ transition strengths.
The present results suggest that 
the ground state has the intermediate structure between cluster and shell-model structures.
In the excited states, well-developed cluster structures are found in both the positive- and negative-parity states. 
Owing to the 3$\alpha$ configurations and the excess neutron motion, various structures appear.

By analyzing the $E2$ transition strengths as well as the GCM amplitudes, 
we assigned members of four positive-parity bands: 
the ground, the triaxial $K^{\pi}=0^{+}_{2}$, the triaxial $K^{\pi}=2^{+}$, and the linear-chain bands.
The $K^{\pi}=0^{+}_{2}$ band and the $K^{\pi}=2^{+}$ side band are constructed from the triaxially deformed intrinsic state, 
in which three $\alpha$ clusters have an almost equilateral-triangular configuration 
and the excess neutrons occupy the $sd$-like orbital.
In the linear-chain band, the excess neutrons distribute 
around two of the three $\alpha$ clusters, which indicates the $^{10}$Be+$\alpha$ correlation. 
This is the first work that suggests the triaxial bands in $^{14}$C. 
For the $3\alpha$ linear-chain band, which has been an attractive topic in C isotopes, 
there have been few theoretical works suggesting the appearance of the linear-chain band in $^{14}$C. 
In the present work, it was found that excess neutrons play two important roles 
in the appearance of the rotational band of the linear-chain structure as follows. 
First, because of the existence of the excess neutrons, 
the energy surfaces for the positive-parity states change to favor the linear-chain structure. 
Second, the excess neutrons are essential for constructing the triaxial bands, 
which occupy the bending $3\alpha$ configurations considerably 
and prevent the linear-chain states from bending to satisfy the orthogonality to each other.
This result suggests an interesting mechanism for the stabilization of the exotic cluster structures 
owing to the excess neutron effects.

For the negative-parity states, because of the largely scattered $E2$ strengths, we could not assign simple band structures.
The origin of these dispersion of the $E2$ strengths is the strong mixing of 
the largely deformed prolate, oblate, and triaxial states in the negative-parity states.
Although the positive-parity linear-chain band with the $^{10}$Be+$\alpha$ correlation was suggested, 
its parity partner with negative parity disappears because of this mixing.
In spite of the largely scattered $E2$ strengths in the negative-parity states,
there are sets of states with significant strong $E2$ transitions. 
These states can be interpreted as members of a quasiband with $^{10}$Be+$\alpha$ cluster structure 
in the negative-parity states of $^{14}$C.

As mentioned earlier, the linear-chain structure shows the $^{10}$Be correlation 
with the asymmetric structure instead of the symmetric configuration of the linear-chain 3$\alpha$+$2n$ structure. 
Further analysis of the $^{10}$Be correlation in the linear-chain structure will be reported in a future article.

The theoretically suggested excited states with $^{10}$Be+$\alpha$ cluster structures 
are the candidates for the excited states which were observed recently in $^{10}$Be+$\alpha$ decays. 
The decay widths and the branching ratios to $^{10}$Be($0^{+}_{1}$)+$\alpha$ and $^{10}$Be($2^{+}_{1}$)+$\alpha$ 
could be helpful for assigning the calculated states to the observed states. 
Theoretical estimation of the partial decay widths of the excited states above the $^{10}$Be+$\alpha$ threshold is also a remaining future problem.

In the present calculations of $^{14}$C, although $\alpha$ clusters are not assumed in the framework, 
three $\alpha$-cluster cores are found to develop in excited states. 
Owing to a variety of 3$\alpha$ configurations and excess neutron motion, 
various interesting structures appear in the $^{14}$C system, 
where the excess neutrons play important roles in the appearance of cluster structures. 
It is a challenging problem to investigate the excited states of further neutron-rich C isotopes such as $^{16}$C 
while focusing on the possibility of cluster states with three $\alpha$-cluster cores with more excess neutrons.

\section*{Acknowledgments}

The computational calculations of this work were performed  by supercomputers in YITP and KEK. 
This work was supported by the YIPQS program in YITP.
It was also supported by the Grant-in-Aid for the Global COE Program 
``The Next Generation of Physics, Spun from Universality and Emergence" 
from the Ministry of Education, Culture, Sports, Science and Technology (MEXT) of Japan,
and Grant-in-Aid for Scientific Research (Nos. 22$\cdot$84, 18540263, and 22540275) from JSPS.

\end{document}